\documentclass[bibyear,a4paper,onecolumn,utf8]{aa}
\usepackage[english]{babel}
\usepackage[utf8]{inputenc}
\usepackage[T1]{fontenc}
\usepackage{aas_macros}
\usepackage{amsmath}
\usepackage{amsthm}
\usepackage{amssymb}
\usepackage{mathtools}
\usepackage{appendix}
\usepackage{rotating}
\usepackage{float}
\usepackage{adjustbox}
\usepackage{color}
\usepackage[normalem]{ulem}
\usepackage{overpic}
\usepackage{url}
\usepackage{subfigure}
\usepackage{dsfont}

\usepackage{multirow}
\usepackage{algorithm}
\usepackage{algorithmic}
\usepackage{textcomp}
\usepackage{gensymb}
\usepackage{graphicx}

\usepackage{txfonts}
\usepackage{natbib}
\bibpunct{(}{)}{;}{a}{}{,} 
\usepackage{xspace}
\usepackage{listings}

\DeclareMathOperator*{\argmin}{arg\,min}
\DeclareMathOperator*{\Id}{Id}

\newcommand{\R}{\mathbb{R}}

\newcommand{\mb} {\mathbf}
\newcommand{\Yb}{\ensuremath{\mb{Y}}}
\newcommand{\Xb}{\ensuremath{\mb{X}}}

\newcommand{\Nb}{\ensuremath{\mb{N}}}
\newcommand{\yb}{\ensuremath{\mb{y}}}
\newcommand{\xb}{\ensuremath{\mb{x}}}
\newcommand{\nb}{\ensuremath{\mb{n}}}
\newcommand{\zb}{\ensuremath{\mb{z}}}
\newcommand{\Hb}{\ensuremath{\mb{H}}}
\newcommand{\Hc}{\ensuremath{\mathcal{H}}}
\newcommand{\Lb}{\ensuremath{\mb{L}}}
\newcommand{\Lc}{\ensuremath{\mathcal{L}}}
\newcommand{\Nc}{\ensuremath{\mathcal{N}_{\boldsymbol{\theta}}}}

\newcommand{\SNR}{\ensuremath{\mathrm{SNR}}}

\definecolor{dkgreen}{rgb}{0,0.6,0}
\definecolor{gray}{rgb}{0.5,0.5,0.5}
\definecolor{mauve}{rgb}{0.58,0,0.82}

\newcommand{\scalefig}[1]{#1\linewidth}

\DeclareTextFontCommand{\emb}{\bfseries\em}

\begin{document}

\title{Deep Learning for space-variant deconvolution in galaxy surveys}
\titlerunning{Deep Learning for galaxy surveys}
\author{F.~Sureau\inst{1}, A.~Lechat \inst{1,2}, J.-L.~Starck\inst{1}}

\authorrunning{F.~Sureau et al.}

\institute{
Laboratoire AIM, CEA, CNRS,
Universit\'e Paris-Saclay, Universit\'e Paris Diderot,
Sorbonne Paris Cit\'e,
F-91191 Gif-sur-Yvette,
France
\and
ONERA - The French Aerospace Lab,
6 chemin de la Vauve aux Granges, BP 80100, FR-91123 PALAISEAU cedex, France
}

\abstract{Deconvolution of large survey images with millions of galaxies requires to develop 
a new generation of methods which can take into account a space variant Point Spread Function (PSF) and 
 have to be at the same time accurate and fast. We investigate in this paper how Deep Learning (DL)  
could be used to perform this task. We employ a U-Net Deep Neural Network (DNN) architecture to learn in a supervised setting parameters adapted for galaxy image processing and study two strategies for deconvolution. The first approach is a post-processing of a mere Tikhonov deconvolution with closed form solution and the second one is an iterative deconvolution framework based on the Alternating Direction Method of Multipliers (ADMM). Our numerical results based on GREAT3 simulations with realistic galaxy images and PSFs show that our two approaches outperforms standard techniques based on convex optimization, whether assessed in galaxy image reconstruction or shape recovery. The approach based on Tikhonov deconvolution leads to the most accurate results except for ellipticity errors at high signal to noise ratio where the ADMM approach performs slightly better, is also more computation-time efficient to process a large number of galaxies, and is therefore recommended in this scenario.}   


\maketitle


\section{Introduction}

Deconvolution of large galaxy survey images requires to take into account spatial-variation of the 
 Point Spread Function (PSF) across the field of view. The PSF field is usually estimated beforehand, via parametric models and simulations as in \citet{Krist2011} or directly estimated from the (noisy) observations of stars in the field of view \citep{Bertin2011,Kuijken2015,Zuntz2018,Mboula2016,Schmitz2019}. 
 Even with the "perfect" knowledge of the PSF, this ill-posed deconvolution problem is challenging, in particular due to the size of the image to process.  \citet{starck:sta00_3} proposed an {\em Object-Oriented Deconvolution}, consisting in 
first detecting  galaxies and then deconvolving each object independently taking into account the PSF at the position of the center of the galaxy (but not taking into account the variation of the PSF field at the galaxy scale). Following this idea, \citet{Farrens2017} introduced a space-variant deconvolution approach for galaxy images, based on two regularization strategies: using either a sparse prior in a transformed domain \citep{Starck2015} or trying to learn unsupervisedly a low-dimensional subspace for galaxy representation using a low-rank prior on the recovered galaxy images. Provided a sufficient number of galaxies are jointly processed (more than 1000) they found that the low-rank approach provided significantly lower ellipticity errors than sparsity, which illustrates the importance of learning adequate representations for galaxies. To go one step further in learning, supervised deep learning techniques taking profit of databases of galaxy images could be employed to learn complex mappings that could regularize our deconvolution problem. Deep convolutional architectures have also proved to be computationally efficient to process large number of images once the model has been learned, and are therefore promising in the context of modern galaxy surveys.\\

{\it Deep Learning and Deconvolution:}
In the recent years, deep learning approaches have been proposed in a large number of inverse problems with high empirical success. Some potential explanations could lie on the expressivity of the deep architectures (e.g. the theoretical works for simple architecture in  \citep{Eldan2015,Safran2017,Petersen2018}) as well as new architectures or new optimization strategies that increased the learning performance (for instance \citet{Kingma2014,Ioffe2015,He2016,Szegedy2016}). Their success also depend on the huge datasets collected in the different applications for training the networks, as well as the increased computing power available to process them.
With the progress made on simulating realistic galaxies (based for instance on real Hubble Space Telescope (HST) images as in \citet{Rowe2015,Mandelbaum2015}), deep learning techniques have therefore the potential to show the same success for deconvolution of galaxy images as in the other applications. Preliminary work have indeed shown that deep neural networks (DNN) can perform well for classical deconvolution of galaxy images \citep{Flamary2017,Schawinski2017}.\\

{\it This paper:} we investigates two different strategies to interface deep learning techniques with space variant deconvolution approaches inspired from convex optimization. In section \ref{sec:deconv}, we review deconvolution techniques based on convex optimization and deep learning schemes. 
The space variant deconvolution is presented in section~\ref{sect_svdeconv} where the two proposed methods are described, the first one using a deep neural network (DNN) for post-processing of a Tikhonov deconvolution and the second one including a DNN trained for denoising in an iterative algorithm derived from convex optimization. The neural network architecture proposed for deconvolution is also presented in this section. The experiment settings  are described in section~\ref{sec:XPDesign} and the results presented in section~\ref{sec:Results}.  We conclude in section \ref{sec:ccl}.
 
\section{Image Deconvolution in the Deep Learning Era}
\label{sec:deconv}
\subsection{Deconvolution before Deep Learning}
The standard deconvolution problem consists in solving the linear inverse problem 
$\mathbf{Y} = \mathbf{H}  \mathbf{X} +  \mathbf{N}$, where $ \mathbf{Y}$ is the observed noisy data, $ \mathbf{X}$ the unknown solution, 
$ \mathbf{H}$ the matrix related to the PSF and $ \mathbf{N}$ is the noise. Images $\mathbf{Y}$,  $\mathbf{X}$ and  $\mathbf{N}$ are represented by a column vector of $n_p$ pixels arranged in lexicographic order, with $n_p$ being the total number of pixels,  and $\mathbf{H}$ is a $n_p \times n_p $ matrix.
State of the art deconvolution techniques typically solve this ill-posed inverse problem (i.e. with no unique and stable solution) through a modeling of the forward problem motivated from physics, and adding regularization penalty term $\mathcal{R}\left(\mathbf{X} \right)$ which can be interpreted as enforcing some constraints on the solution. It leads to minimize:
\begin{equation}
  \label{eq:Regularization}
  \argmin\limits_{\mathbf{X}} \frac{1}{2}|| \mathbf{Y} - \mathbf{H}  \mathbf{X}  ||^2_F + \mathcal{R}\left(\mathbf{X} \right),
\end{equation}
where  $||\cdot||_F$ is the Frobenius norm.  The most simple (and historic) regularization corresponding  is the Tikhonov regularization \citep{Tikhonov1977,Hunt1972,Twomey1963}, where $\mathcal{R}\left(\mathbf{X} \right)
$ is a  quadratic term,  $\mathcal{R}\left(\mathbf{X} \right) =  \frac{\lambda}{2} ||\mathbf{L} \Xb||^2_F$. 
The closed-form solution of this  inverse problem is given by:
\begin{equation}
  \label{eq:TikhonovSolution}
\tilde{\mathbf{X}} = \left( \mathbf{H}^T  \mathbf{H} +\lambda  \mathbf{L}^T \mathbf{L} \right)^{-1}  \mathbf{H}^T \mathbf{Y}
 \end{equation}
which involves the Tikhonov linear filter $\left(\Hb^T \Hb + \lambda  \Lb^T  \Lb \right)^{-1} \Hb^T$. The simplest version is when  $\Lb =\boldsymbol{\Id}$, which penalizes solutions with high energy. When  the PSF is space invariant, the matrix $ \mathbf{H}$ is block circulant and the inverse problem can then be written as a simple convolution product. It is also easy to see that the Wiener deconvolution corresponds to a specific case of the Tikhonov filter. See \citet{ima:bertero98} for more details.

 \begin{figure}[ht]\centering
  \includegraphics[width=\scalefig{0.95}]{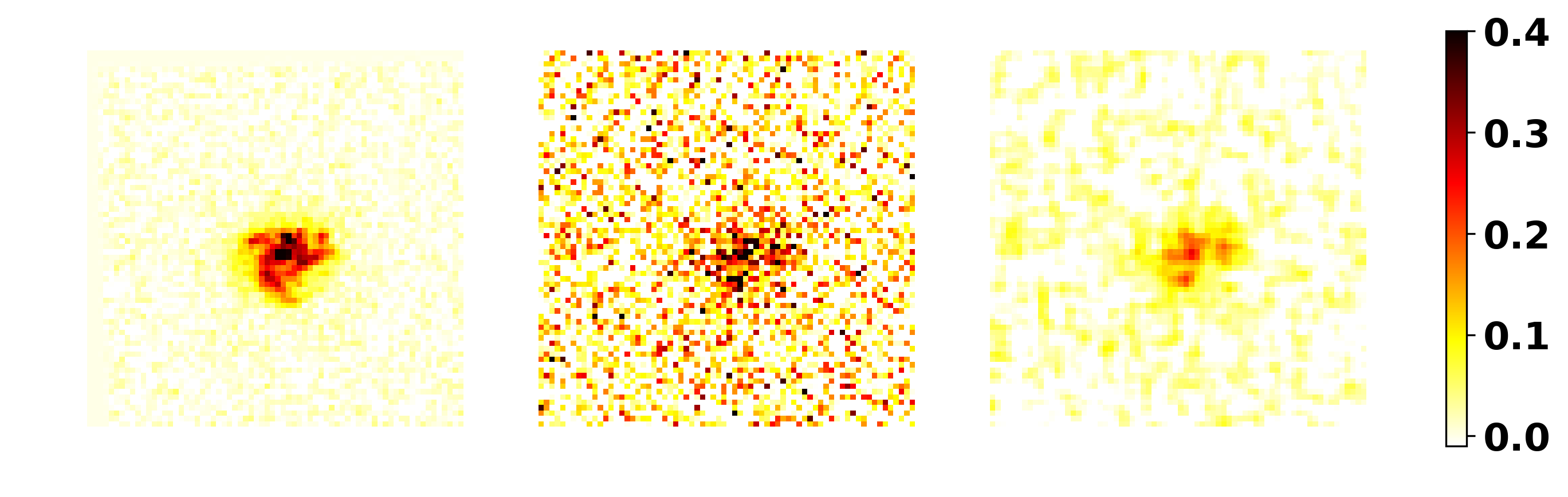}
  \caption{Deconvolution with Tikhonov regularization.From left to right:galaxy image from HST used for the  simulation, observed galaxy at $\SNR=20$ (see below for our definition of SNR), deconvolved image computed from Eq.~\ref{eq:TikhonovSolution}.
  }
  \label{fig:TikhonovExample}
\end{figure}
This rather crude deconvolution is illustrated in Fig.~\ref{fig:TikhonovExample} in a low signal to noise ratio (SNR) scenario, displaying both oversmoothing of the galaxy image, loss of energy in the recovered galaxy and the presence of coloured noise due to the inverse filter. 

Most advanced methods are non linear and generally involves iterative algorithms. There is a vast litterature in image processing on advanced regularization techniques applied to deconvolution: adding some prior information on $\Xb$ in a Bayesian paradigm \citep{BioucasDias2006,Krishnan2009, Orieux2010} or assuming $\Xb$ to belong to some classes of images to recover (e.g. using total variation regularization \citep{Oliveira2009,Cai2010}, sparsity in fixed representations \citep{Starck2003,Pesquet2009,Pustelnik2016} or learnt via dictionary learning \citep{Mairal2008, Lou2011, Jia2011}), by constraining the solution to belong to some convex subsets (such as ensuring the final galaxy image to be positive). 

For instance, a very efficient approach used for galaxy image deconvolution is based on sparse recovery which consists in minimizing:
\begin{equation}
    \begin{aligned}
        & \argmin\limits_{\mathbf{X}} \frac{1}{2}\|\mathbf{Y}-\mathbf{H}\mathbf{X} \|_2^2 + \lambda\|\boldsymbol{\Phi}^T \mathbf{X}\|_1
    \end{aligned}
\label{eq:sparsel2minpos}
\end{equation}   
where $\boldsymbol{\Phi}$ is a matrix related to a fixed transform (i.e. Fouriers, wavelet, curvelets, etc) or that can be learned 
from the data or a training data set  \citep{starck:book15}. The $\ell_1$ norm in the regularisation term is known 
to reinforce the sparsity of the solution, see \citet{starck:book15} for a  review on sparsity.  Sparsity was found extremely efficient for different 
inverse problems in astrophysics such as Cosmic Microwave Background (CMB) estimation \citep{PR1_WPR1}, compact sources estimation in CMB missions \citep{Sureau2014}, weak lensing map recovery \citep{glimpse2016} 
or radio-interferomety image reconstruction \citep{starck:garsden2015}. We will compare in this work our deconvolution techniques with such sparse deconvolution approach.

Iterative convex optimization techniques have been devised to solve Eq.\ref{eq:sparsel2minpos} (see for instance \citet{Beck2009,Zibulevsky2010,Combettes2011,Chambolle2011,Afonso2011,Condat2013,Combettes2014}), with well-studied convergence properties, but with a high computing cost when using adaptive representation for galaxies. This problem opens the way to a new generation of methods. 

\subsection{Toward Deep Learning}
Recently deep learning techniques have been proposed to solve inverse problems by taking benefit of the dataset collected and/or the advances in simulations, including for deconvolving galaxy images. These  approaches have proved to be able to learn complex mappings in the supervised setting, and to be computationally efficient once the model has been learned. We review here, without being exhaustive, some recent work on deconvolution using DNNs. We have identified three different 
strategies for using DNN in a deconvolution problem:
\begin{itemize}
\item{\bf Learning the inverse: } the inverse convolution filter can be directly approximated using convolutional neural networks \citep{Xu2014,Schuler2016}. In our application with space-variant deconvolution and known kernels, such complicated blind deconvolution is clearly not necessary and would require a large amount of data to try learning information already provided by the physics included in the forward model. 
\item{\bf Post-processing of a regularized deconvolution:} In the early years of using sparsity for deconvolution  a two steps  approach was proposed,   consisting in first applying a simple linear deconvolution such as using the pseudo-inverse or the Tikhonov filter, letting noise entering in the solution, and then in the second step applying a sparse denoising (see the wavelet-vaguelette decomposition \citep{Donoho1995,Kalifa2003}, more general regularization \citep{Guerrero2006}, or the ForWaRD method  \citep{Neelamani2004}). Similarly, the second step have been replaced by denoising/removing artefacts using a multi-layer perceptron \citep{Schuler2013}, or more recently using U-Nets  \citep{Jin2017}. CNNs are well adapted to this tasks, since the form of a CNN mimics unrolled iterative approaches when the forward model is a convolution.  
In another application, convolutional networks such as deep convolutional framelets have also been applied to remove artefacts from reconstructed CT images \citep{Ye2018}. One advantage of such decoupling approach is the ability to process quickly a large amount of data when the network has been learnt, if the  deconvolution chosen has closed-form expression. 
\item{\bf Iterative Deep Learning: } the third strategy uses iterative approaches often derived from convex optimization coupled with deep learning networks. Several schemes have been devised to solve generic inverse problems. The first option, called unrolling or unfolding (see \citet{Monga2019} for a detailed review), is to mimic a few iterations of an iterative algorithm with DNNs so as to capture in the learning phase the impact of 1) the prior \citep{Mardani2017}, 2) the hyperparameters \citep{Mardani2017,Adler2017,Adler2018,Bertocchi2019}, 3) the updating step of a gradient descent \citep{Adler2017} or 4) the whole update process \citep{Gregor2010,Adler2018,Mardani2018}. Such approaches allow fast approximation of iterative algorithms \citep{Gregor2010}, better hyperparameter selection \citep{Bertocchi2019}  and/or provide in a supervised way new algorithms \citep{Adler2017,Mardani2018,Adler2018} better adapted to process specific data set. This approach has noticeably been used recently for blind deconvolution \citep{Li2019}.
Finally, an alternative is to use iterative proximal algorithms from convex optimization (for instance in the framework of the alternating direction method of multiplier plug\&play (ADMM PnP) \citep{Venkatakrishnan2013,Sreehari2016, Chan2016}, or regularization by denoising \citep{Romano2017,Reehorst2018}),  where the proximity operator related to the prior is replaced by a DNN \citep{Meinhardt2017,Bigdeli2017,Gupta2018} or a series of DNN trained in different denoising settings as in \citet{Zhang2017}. 
\end{itemize}
The last two strategies are therefore more adapted to our targeted problem, and in the following we will investigate how they could be applied and how they perform compared to state-of-the art methods in space-variant deconvolution of galaxy images. 

\subsection{Discussion relative to Deep Deconvolution and Sparsity}
It is interesting to notice that connections exist between sparse recovery methodology and DNN:
\begin{itemize}
\item{\textit{Learning Invariants}}: the first features learnt in convolutive deep neural networks correspond typically to edges at particular orientation and location in the images \citep{Lecun2015}, which is also what the wavelet transforms extract at different scales. 
Similar observations were noted for features learnt with a CNN in the context of cosmological parameter estimations from weak-lensing convergence maps \citep{Ribli2018}. As well, understanding mathematically how the architecture of such networks captures progressively powerful invariants can be approached via wavelets and their use in the wavelet scattering transform \citep{Mallat2016}.
\item{\textit{Learned proximal operator}:} \citet{Meinhardt2017} has shown that using a denoising neural network instead of a proximal operator (e.g. soft-thresholding in wavelet space in sparse recovery) during the minimisation iterations improves the deconvolution performance. They also claim that the noise level used to train the neural network behave like the regularisation parameter in sparse recovery. The convergence of the algorithm is not guaranteed anymore, but they observed experimentally that their algorithms stabilize and they expressed their fixed-points.
\item{\textit{expanding path} and \textit{contracting path}:} the U-nets two parts are very similar to synthesis and analysis concepts in sparse representations. This has motivated the use of wavelets to implement in the U-net average pooling and unpooling in the expanding path \citep{ye2018deep,han2018framing}. Some other connection can be made with soft-Autoencoder in \citet{fan2018soft} introducing a pair of ReLU units emulating soft-thresholding, accentuating the comparison with cascade wavelet shrinkage systems.   
\end{itemize}
Therefore, we observe exchanges between the two fields, in particular for U-Net architectures, with however significant differences such as the construction of a very rich dictionary in U-nets that is possible through the use of a large training data set, as well as non-linearities at every layer essential to capture invariants in the learning phase.

\section{Image Deconvolution with Space Variant PSF}
\label{sect_svdeconv}
\subsection{Introduction}
In the case of a space-variant deconvolution problem, we can write the same deconvolution equation as before, $\Yb = \Hb \Xb + \Nb$, but  $\Hb$ is not block circular anymore, and manipulating such a huge matrix is not possible in practice. 
As in \citet{Farrens2017}, we consider instead an {\em Object-Oriented Deconvolution}, by first detecting $n_g$ galaxies with  $n_p$ pixels each and then deconvolving independently each object using the PSF at the position of the center of the galaxy. We use the following definitions: 
 the observations of $n_g$ galaxies with $n_p$ pixels are collected in $\Yb\in\R^{n_p\times n_g}$ (as before, each galaxy being represented by a column vector arranged in lexicographic order), the galaxy images to recover are similarly collected $\Xb\in\R^{n_p\times n_g}=[\xb_\mb{i}]_{i=1..n_g}$ and the convolution operator with the different kernels is noted $\Hc$. It corresponds to applying in parallel a convolution matrix $\Hb_\mb{i}$ to a galaxy $\xb_\mb{i}$ ($\Hb_\mb{i}$ being typically a block circulant matrix with circulant block after zero padding which we perform on the images \citep{Andrews77}). Noise is noted $\Nb\in\R^{n_p\times n_g}$ as before and is assumed to be additive white gaussian noise. With these definitions, we now have 
\begin{equation}
  \label{eq:forwardModel}
  \Yb = \Hc(\Xb)+\Nb
\end{equation}
or more precisely
\begin{equation}
  \label{eq:forwardModelDevelop}
  \left\{\yb_\mb{i} = \Hb_\mb{i}\xb_\mb{i}+\nb_\mb{i}\right\}_{i=1..n_g}
,\end{equation}
for block circulant $\left\{\Hb_\mb{i}\right\}_{i=1..n_g}$, which illustrates that we consider multiple local space-invariant convolutions in our model (ignoring the very small variations of the PSF at the scale of the galaxy as done in practice \citep{Kuijken2015,Mandelbaum2015,Zuntz2018}). 
The deconvolution problem of finding $\Xb$ knowing $\Yb$ and $\Hc$  is therefore considered as a series of independent ill-posed inverse problems. To avoid having multiple solutions (due to a non trivial null space of $\left\{\Hb_\mb{i}\right\}_{i=1..n_g}$) or an unstable solution (bad conditioning of these matrices), we need to regularize the problem as in standard deconvolution approaches developed for space-invariant convolutions. This amounts to solve the following inverse problem:
\begin{equation}
  \label{eq:SVRegularization}
  \argmin\limits_{\Xb} \frac{1}{2}||\Yb- \Hc(\Xb)||^2_F + \mathcal{R}\left(\Xb\right)
\end{equation}
and in general we will choose separable regularizers so that we can handle in parallel the different deconvolution problems:
\begin{equation}
  \label{eq:RegularizationDevelop}
  \left\{\argmin\limits_{\xb_\mb{i}} \frac{1}{2}||\yb_\mb{i}-  \Hb_\mb{i}\xb_\mb{i}||^2_2 + \mathcal{R}\left(\xb_\mb{i}\right)\right\}_{i=1..n_g}
\end{equation}

\citet{Farrens2017} proposes two methods to perform this deconvolution:
\begin{itemize}
\item{\bf Sparse prior:} each galaxy is supposed to be sparse in the wavelet domain, leading to minimize 
\begin{equation}
    \begin{aligned}
        & \underset{\mathbf{X}}{\text{argmin}}
        & \frac{1}{2}\|\mathbf{Y}-\mathcal{H}(\mathbf{X})\|_2^2 + 
        \|\mathbf{W}^{(k)}\odot\Phi(\mathbf{X})\|_1
        & & \text{s.t.}
        & & \mathbf{X} \ge 0
    \end{aligned}
    \label{eq:rw_l1}
\end{equation}
with $\mathbf{W}^{(k)}$ a weighting matrix.
\item{\bf Low rank prior:} In the above method, each galaxy is deconvolved independently from the others. 
As there are many similarities between galaxy images,  \citet{Farrens2017} proposes a joint restoration process, where the matrix
$\mathbf{X}$ has a low rank. This is enforced by adding a nuclear norm penalization instead of the sparse regularization, as follows:
\begin{equation}
    \begin{aligned}
        & \underset{\mathbf{X}}{\text{argmin}}
        & \frac{1}{2}\|\mathbf{Y}-\mathcal{H}(\mathbf{X})\|_2^2 + \lambda\|\mathbf{X}\|_*
        & & \text{s.t.}
        & & \mathbf{X} \ge 0
    \end{aligned}
    \label{eq:lowrl2minpos}
\end{equation}  
where  
$ \|\mathbf{X}\|_* = \sum_{k} \sigma_k(\mathbf{X})$, $\sigma_k(\mathbf{X})$ denoting the $k^{\text{th}}$ largest singular value of $\mathbf{X}$.  
\end{itemize}
It was shown that the second approach outperforms sparsity techniques as soon as the number of galaxies in the field is larger than 1000 \citep{Farrens2017}.  

\subsection{Neural Network architectures}
\label{subsec:architecture}
DNN allows us to extend the previous low rank minimisation, by taking profit of existing databases and learning more features from the data in a supervised way, compared to what we could do with the simple SVD used for nuclear norm penalization. 
The choice of network architecture is crucial for performance. We have identified three different features we believe important for our application: 1) the forward model and the task implies that the network should be translation equivariant, 2) the model should include some multi-scale processing based on the fact that we should be able to capture distant correlations, and 3) the model should minimize the number of trainable parameters for a given performance, so as to be efficient (lower GPU memory consumption) which is also important to ease the learning. 
Hopefully these objectives are not contradictory: the first consideration leads to the use of convolutional layers, while the second implies a structure such as the U-Net \citep{Ronneberger2015} already used to solve inverse problems \citep{Jin2017} or the deep convolutional framelets \citep{Ye2018}. But because such architectures allow to increase rapidly the receptive field in the layers along the network, they can compete with a smaller number of parameters against CNNs having a larger number of layers and therefore more parameters.

\begin{figure}[ht]\centering
  \includegraphics[width=0.5\linewidth]{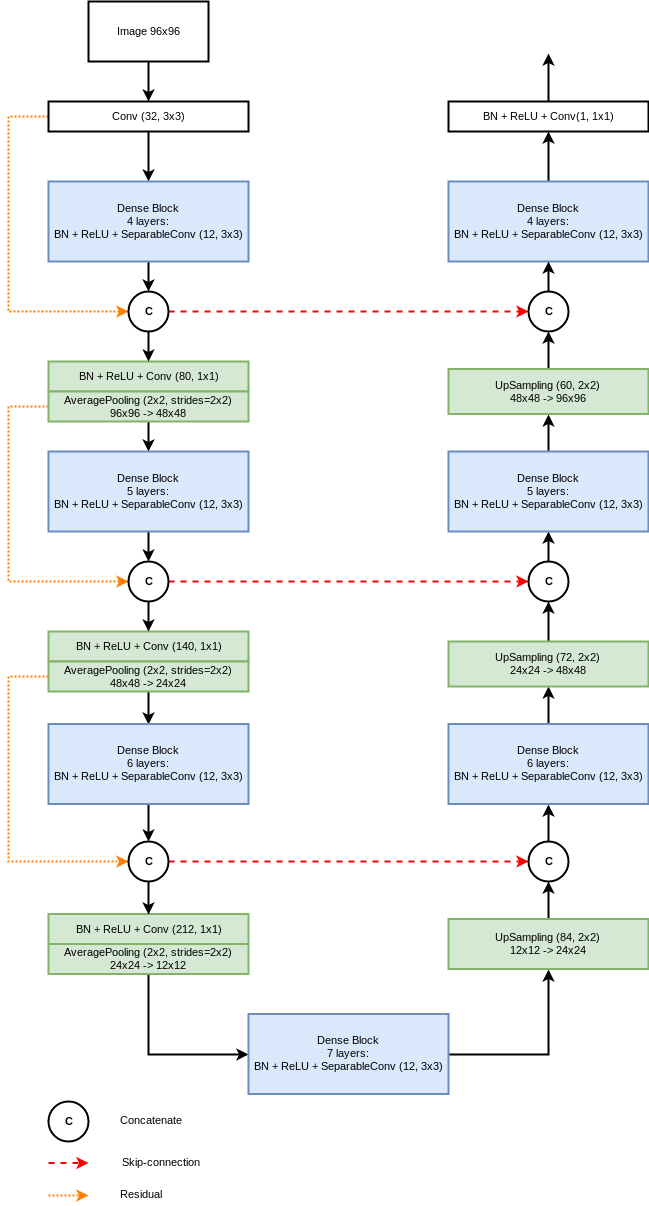}
  \caption{DNN model used in this work. The global architecture is a U-Net, with small modifications for performance and to limit the number of model parameters.}
  \label{fig:DenseNet}
\end{figure}

We therefore have selected  a global U-Net structure as in \citep{Jin2017}, but including the following modifications:
\begin{itemize}
\item{\bf 2D separable convolutions:} we replace 2D convolutions by 2D separable convolutions \citep{Chollet2016}. The separable convolutions allow to limit the number of parameters in the model by assuming that spatial correlations and  correlations across feature maps can be independently captured. Their use have already lead to outperform architectures with non-separable convolution with a larger number of parameters \citep{Chollet2016}. 
\item{\bf Dense blocks:} we changed the convolutional layers at each "scale" by using dense blocks \citep{Huang2017}. 
Dense blocks also allow to reduce the number of parameters, by propagating through concatenation all prior feature maps to the input of the current layer.  This was claimed to enable feature reuse, preservation of information, and to limit vanishing gradients in the learning. 
\item{\bf Average-pooling:} we change the pooling step: we have observed that max-pooling lead to over-segmentation of our final estimates, which is alleviated by the use of average pooling.
\item{\bf Skip connection:} we removed the skip connection between the input and the output layers introduced by \citep{Jin2017} which proved to be detrimental to the performance of the network, especially at low SNR. Note that the dense blocks may have also better preserved the flow of relevant information and limited the interest of using residual learning. 
\end{itemize}
The two first modification limit significantly the number of parameters per "scale" of the U-Net, and potentially allow for more scales to be used for a given budget of number of trainable parameters. Our network, we name "XDense U-Net", is displayed in Fig.~\ref{fig:DenseNet}. The following describes how to use such networks in two different ways in order to perform the space variant deconvolution.

\subsection{Tikhonet: Tikhonov deconvolution post-processed by a Deep Neural Network}
\label{subsec:Tikhonet}
The Tikhonov solution for the space variance variant PSF deconvolution is:
\begin{equation}
  \label{eq:TikhonovSVRegularization}
  \argmin\limits_{\Xb} \frac{1}{2}||\Yb- \Hc(\Xb)||^2_F +  ||\Lc(\Xb)||^2_F
\end{equation}
where $\Lc$ is similarly built as $\Hc$.
The closed-form solution of this linear inverse problem is given by:
\begin{equation}
  \label{eq:SVTikhonovSolution}
 \left\{\tilde{\xb_\mb{i}} = \left(\Hb^T_\mb{i}\Hb_\mb{i}+\lambda_i\Lb^T_\mb{i}\Lb_\mb{i} \right)^{-1} \Hb_\mb{i}^T \yb_\mb{i}\right\}_{i=1..n_g}
 \end{equation}
which involves for each galaxy a different Tikhonov filter $\left(\Hb^T_\mb{i}\Hb_\mb{i}+\lambda_i\Lb^T_\mb{i}\Lb_\mb{i} \right)^{-1} \Hb_\mb{i}^T$. In this work, we chose $\Lb_\mb{i}=\boldsymbol{\Id}$  and the regularization parameter $\lambda_i$ is different for each galaxy, depending on its SNR (see \ref{sec:proposedDL} for more details).
The final estimate is then only: 
\begin{equation}
  \label{eq:TikhoPredict}
\left\{\hat{\xb_\mb{i}}=\Nc(\tilde{\xb_\mb{i}})\right\}_{i=1..n_g},
\end{equation}
where the neural network predictions based on its parameters $\boldsymbol{\theta}$ for some inputs $\Yb$ are written as $\Nc(\Yb)$. 

The success of the first approach therefore lies on the supervised learning of the mapping between the Tikhonov deconvolution of Eq.~\eqref{eq:SVTikhonovSolution} and the targeted galaxy.

We call this two-step approach {\bf "Tikhonet"} and the rather simple training process is described in Algorithm~\ref{Algo:Tikhonet}.

\begin{algorithm}
  \caption{DNN training in the Tikhonet approach \label{Algo:Tikhonet}}
  \begin{algorithmic}[1]
  \STATE \textbf{Initialization}: Prepare noise-free training set, choose noise parameters (SNR range) and validation set.   Choose architecture for network $\mathcal{N}$,  learning parameters (optimizer and its parameters, batch size $B$ and number of batches $n_{batch}$, number of epochs $n_{epoch}$) and cost function to minimize (here mean squared error).
  \FOR[\textbf{Loop over epochs}]{$n=1$ to $n_{epoch}$}
  	\FOR[\textbf{Loop over batches}]{$b=1$ to $n_{batch}$}
  		\FOR[\textbf{Loop over galaxies in batch}]{$i=1$ to $B$}
  			\STATE Add random noise to obtain a realization in the SNR range chosen
  			\STATE Compute the Tikhonov solution  $\tilde{\xb_\mb{i}}$ using Eq.~\ref{eq:TikhonovSolution}
  		\ENDFOR
  		\STATE Predict $\hat{\xb_\mb{i}}$ (Eq.~\ref{eq:TikhoPredict}) and update network parameters $\boldsymbol{\theta}$  according to the cost function.
  	\ENDFOR
  \ENDFOR
  \RETURN  $\Nc$
  \end{algorithmic}
\end{algorithm}

\subsection{ADMMnet: Deep neural networks as constraint in ADMM plug-and-play}
\label{subsec:ADMMnet}
The second approach we investigated is using the ADMM PnP framework with a DNN. 

ADMM is an augmented lagrangian technique developed to solve convex problems under linear equality constraints (see for instance \citep{Boyd2010}). It operates by decomposing the minimization problem into sub-problems solved sequentially. One iteration consists in first solving a minimization problem typically involving the data fidelity term, then solving a second minimization problem involving the regularization term, and finishing by an update of the dual variable. 

It has previously been noted \citep{Venkatakrishnan2013,Sreehari2016, Chan2016} that the first two sub-steps can be interpreted as an inversion step followed by a denoising step coupled via the augmented lagrangian term and the dual variable. These authors suggested to use such ADMM structure with non-linear denoisers in the second step in an approach dubbed ADMM PnP, which recent work has proposed to implement via DNNs \citep{Meinhardt2017}. 

In the following, we adopt such iterative approach based on the ADMM PnP because 1) it separates the inversion step and the use of the DNN, offering  flexibility to add extra convex constraints in the cost function that can be handled with convex optimization 2) it alleviates the cost of learning by focusing essentially on learning a denoiser or a projector - less networks, less parameters to learn jointly compared to unfolding approaches where each iteration corresponds to a different network 3) by iterating between the steps, the output of the network is propagated to the forward model to be compared with the observations, avoiding large discrepancies, contrary to the Tikhonet approach where the output of the network is not used in a likelihood.

The training of the network $\Nc$ in this case is similar to Algorithm~\ref{Algo:Tikhonet}, except that the noise-free training set is composed of noise-free target images instead of noise-free convolved images, and the noise added has constant standard deviation. Then the algorithm for deconvolving a galaxy is presented in Algo.~\ref{Algo:ADMMPnP} and is derived from \cite{Chan2016}. The application of the network is here illustrated in red. We call this approach {\bf "ADMMnet"}. The first step consists in solving the following regularized deconvolution problem at iteration $k$ using the accelerated iterative convex algorithm FISTA \citep{Beck2009}:
\begin{equation}
  \label{eq:RegularizationADMM}
  \left\{\argmin\limits_{\xb_\mb{i}} \frac{1}{2\sigma^2}||\yb_\mb{i}-  \Hb_\mb{i}\xb_\mb{i}||^2_2 + \iota_\mathcal{C}(\xb_\mb{i}) + \frac{\rho}{2}||\xb_\mb{i}-\zb^{(k)}_\mb{i}+\boldsymbol{\mu}^{(k)}||^2_2 \right\}_{i=1..n_g}
\end{equation}
where $\iota_\mathcal{C}$ is the characteristic function of the  non-negative orthant, to enforce the non-negativity of the solution. The DNN used in the second step is used as an analogy with a denoiser (or as a projector),as presented above. The last step controls the augmented lagrangian parameter, and ensure that this parameter is increased when the optimization parameters are not sufficiently changing. This continuation scheme is also important, as noted in \citet{Chan2016}, as increasing progressively the influence of the augmented lagrangian parameter ensures stabilization of the algorithm.

Note that of course there is no convergence guarantee of such scheme and that contrary to the convex case the augmented lagrangian parameter $\rho$ is expected to impact the solution.
 
Finally, because the target galaxy is obtained after re-convolution with a target PSF to avoid aliasing (see section \ref{sec:XPDesign}), we also re-convolve the ADMMnet solution with this target PSF to obtain our final estimate.

\begin{algorithm}
  \caption{Proposed ADMM Deep Plug\&Play for deconvolution of a galaxy image\label{Algo:ADMMPnP}}
  \begin{algorithmic}[1]
  \STATE \textbf{Initialize}:set $\rho_0,\rho_{max},\eta\in[0,1),\gamma>1,\Delta_0=0$,$\mathbf{X}^{(0)}=\mb{0},\mathbf{Z}^{(0)}=\mb{0},\boldsymbol{\mu}^{(0)}=\mb{0}, \epsilon$ 
  \FOR[\textbf{Main Loop}]{$k=0$ to $N_{it}$}
  \STATE \textbf{Deconvolution Sub-Problem}: $\mathbf{X}^{(k+1)}=FISTA(\textbf{Y},\mathbf{X}^{(k)},\mathbf{Z}^{(k)},\boldsymbol{\mu}^{(k)},\rho_k)$  
  \STATE \textbf{"Denoising" Sub-Problem}: $\mathbf{Z}^{(k+1)}=\textcolor{red}{\Nc}\left(\mathbf{X}^{(k+1)}+\boldsymbol{\mu}^{(k)}\right)$ 
  \STATE \textbf{Lagrange Multiplier Update}: $\boldsymbol{\mu}^{(k+1)}=\boldsymbol{\mu}^{(k)}+\left(\mathbf{X}^{(k+1)}-\mathbf{Z}^{(k+1)}\right)$
  \STATE $\Delta_{k+1}=\frac{1}{\sqrt{n}}\left(||\mathbf{X}^{(k+1)}-\mathbf{X}^{(k)}||_2+||\mathbf{Z}^{(k+1)}-\mathbf{Z}^{(k)}||_2+||\boldsymbol{\mu}^{(k+1)}-\boldsymbol{\mu}^{(k)}||_2\right)$
  \IF{$\Delta_{k+1}\geq \eta \Delta_k$ \AND  $\rho_{k+1}\leq \rho_{max}$} 
      \STATE  $\rho_{k+1}=\gamma\rho_k$
  \ELSE
      \STATE  $\rho_{k+1}=\rho_k$
  \ENDIF
   \IF{$\|\mathbf{Z}^{(k+1)}-\mathbf{X}^{(k+1)}\|_2<\epsilon$}
   		\STATE \textbf{stop}
   \ENDIF 

\ENDFOR
\RETURN $\left\{\mathbf{X}^{(k+1)}\right\}$
  \end{algorithmic}
\end{algorithm}

\subsection{Implementation and choice of parameters for Network architecture}
\label{sec:proposedDL}

We  describe here our practical choices for the  implementation of the algorithms.
For the Tikhonet, the hyperparameter $\lambda_i$ that controls the balance in between the data fidelity term and the quadratic regularization in Eq.~\ref{eq:SVTikhonovSolution} needs  to be set for each galaxy. This can be done either manually by selecting an optimal value as a function of an estimate of the SNR,  or by using automated procedures such as generalized cross-validation (GCV) \citep{Golub1979}, the L-curve methode \citep{Hansen1993}, the Morozov discrepancy principle \citep{Engl1996}, various Stein Unbiased Risk Estimate (SURE) minimization \citep{Eldar2009,Pesquet2009,Deledalle2014}, or using a  hierarchical Bayesian framework \citep{Orieux2010,Pereyra2015}. We compared these approach, and report the results obtained by the SURE prediction risk minimization which lead to the best results with the GCV approach.
 
For the ADMM, the  parameters $\rho_{0}$, $\rho_{max}$, $\eta$, $\epsilon$ and $\gamma$ have been selected manually, as a balance between  stabilizing quickly the algorithm (in particular high $\rho$) and favouring the minimization of the data fidelity term in the first steps (low $\rho$). We investigated in particular the choice of $\rho_{0}$ which illustrate how the continuation scheme impacts the solution.

The DNNs were coded in Keras\footnote{\url{https://keras.io}} with Tensorflow \footnote{\url{https://www.tensorflow.org}} as backend. For the proposed XDense U-Net, 4 scales were selected with an increasing number of layers for each scale (to capture distant correlations). Each separable convolution was composed of $3\times 3$ spatial filters and a growth factor of 12 was selected for the dense blocks. The total number of trainable parameters was 184301. 
We also implemented a "classical" U-Net to test the efficiency of the proposed XDense U-Net architecture. For this U-Net, we choose 3 scales with 2 layers per scale and 20 feature maps per layer in the first scale, to end up with 206381 trainable parameters ($12\%$ more than the XDense U-Net implementation). 
In both networks we used batch normalization and rectified linear units for the activation. We also tested for our proposed approach weighted sigmoid activations (or swish in \citet{Elfwing2018}) which seems to slightly improve the results but at the cost of increasing the computational burden and therefore we did not use them in the following results.

In the training phase, we use 20 epochs, a batch size of 32 and the Adam optimizer was selected (we keep the default parameters) to minimize the mean squared error (MSE) cost function. After each epoch, we save the network parameters only if they improve the MSE on the validation set.

\section{Experiments}
\label{sec:XPDesign}
In this section, we describe how we generated the simulations used for learning networks and testing our deconvolution schemes, as well as the criteria we will use to compare the different deconvolution techniques.

\subsection{Dataset generation}
\begin{figure}[ht]\centering
  \includegraphics[width=\scalefig{0.75}]{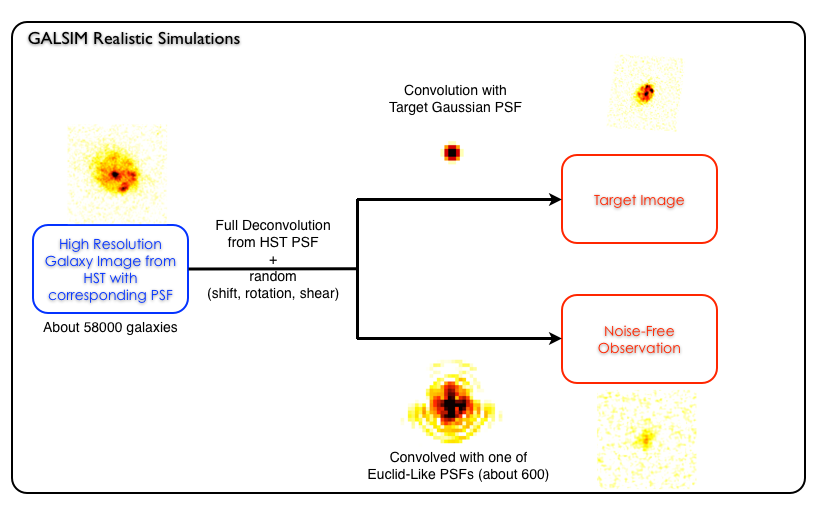}
  \caption{Set up for a GalSim simulated realistic galaxy. In the upper branch we obtain the targeted galaxy. In the lower branch, we simulate the corresponding Euclid-like observed galaxy. Note that in these figures, a log-scale was adopted for the PSFs to illustrate its complicated structure.  }
  \label{fig:Galsim}
\end{figure}

We use GalSim\footnote{\url{https://github.com/GalSim-developers/GalSim}} \citep{Rowe2015} to generate realistic images of galaxies for training our networks and testing our deconvolution approaches.  We essentially follow the approach used  in GREAT3 \citep{Mandelbaum2014} to generate the realistic space branch from  high resolution HST images, but choosing the PSFs in a set of 600 Euclid-like PSFs (the same as in \citet{Farrens2017}). The process is illustrated in Fig.~\ref{fig:Galsim}. 

A HST galaxy is randomly selected from the set of about 58000 galaxies used in the GREAT3 challenge, deconvolved with its PSF, and random shift (taken from a uniform distribution in $[-1,1]$ pixel), rotation and shear are applied. The same cut in SNR is performed  as in GREAT3  \citep{Mandelbaum2014} , so as to obtain a realistic set of galaxies that would be observed in a SNR range $[20,100]$ when the noise level is as in GREAT3.  In this work we use the same definition of SNR as in this challenge:
\begin{equation}
  \label{eq:SNR}
\SNR\left(\Xb_\mb{i}\right)=\frac{||\Xb_\mb{i}||_2}{\sigma}
 \end{equation}
where $\sigma$ is the standard deviation of the noise. This SNR corresponds to an optimistic SNR for detection when the galaxy profile $\Xb_\mb{i}$ is known. In other (experimental) definitions, the minimal SNR is indeed closer to 10, similarly to what is usually considered in weak lensing studies  \citep{Mandelbaum2014}.

If the cut in SNR is passed,  to obtain the target image in a $96\times96$ grid with pixel size $0.05''$,  we first convolve the HST deconvolved galaxy image with a Gaussian PSF with $FWHM=0.07''$ to ensure no aliasing occurs after the subsampling. To simulate the observed galaxy without extra noise, we convolve the HST deconvolved image with a PSF randomly selected among about 600 Euclid-like PSFs  (the same set as used in \citet{Farrens2017}). Note that the same galaxy rotated by $90\degree$ is also simulated as in GREAT3.

Because we use as inputs real HST galaxies, noise from HST images propagate to our target and observed images, and is coloured by the deconvolution/reconvolution process. We did not want to denoise the original galaxy images to avoid losing substructures in the target images (and making them less "realistic"), and as this noise level is lower than the noise added in our simulations we expect it to change marginally our results - and not the ranking of methods. 

This process is repeated so that we end up with about 210000 simulated observed galaxies and their corresponding target. For the learning, 190000 galaxies are employed, and 10000 for the validation set. The extra 10000 are used for testing our approaches.\\

\begin{figure}[ht]\centering
  \includegraphics[width=\scalefig{0.95}]{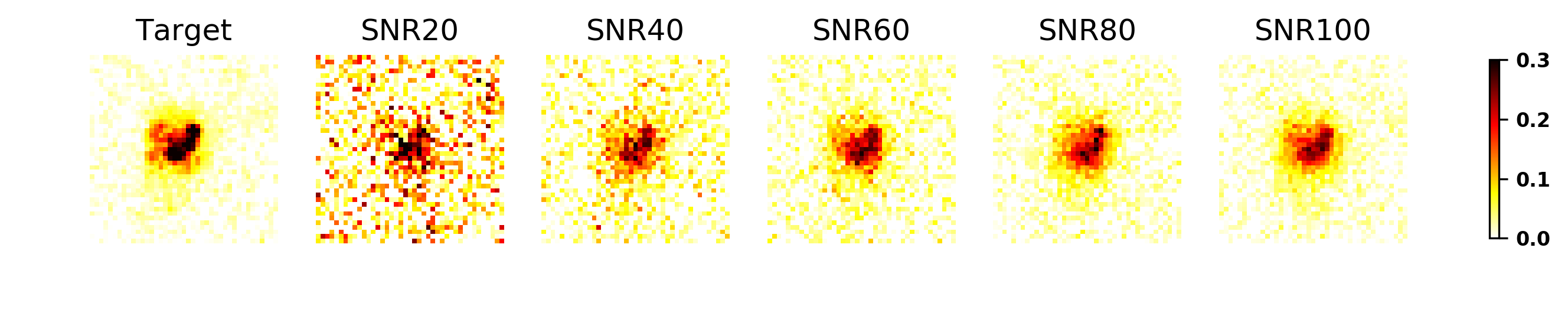}
  \caption{Range of SNR used for the training and for testing in the simulations. From left to right: targeted galaxy image, then observed convolved images at increasing SNR.  In our definition, $\SNR=20$ is barely at the galaxy detection limit, while at $\SNR=100$ galaxy substructures can be visualized.}
  \label{fig:SNRRange}
\end{figure}

In the learning phase,  additive white Gaussian noise is added to the galaxy batches with standard deviation chosen so as to obtain a galaxy in a prescribed SNR range. For the Tikhonet, we choose randomly for each galaxy in the batch a $\SNR$ in the range $[20,100]$, which corresponds to selecting galaxies from the limit of detection to galaxies with observable substructures, as illustrated in Fig~\ref{fig:SNRRange}. For the ADMMnet, we learn a denoising network for a constant noise standard deviation of $\sigma=0.04$ (same level as in GREAT3).

\begin{figure}[ht]\centering
  \includegraphics[width=\scalefig{0.45}]{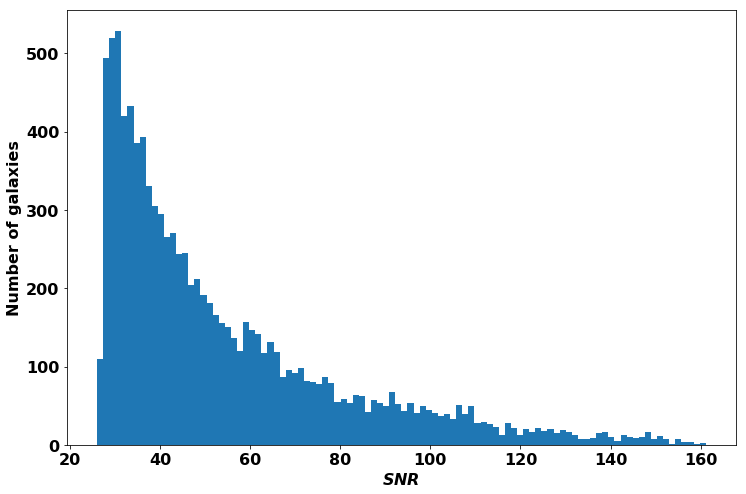}
  \caption{Distribution of SNR of simulated galaxies for constant noise simulations ($\sigma=0.04$). The peak of the distribution is at about $\SNR=30$, and the mean SNR is $\SNR=54$.}
  \label{fig:histoSNR}
\end{figure}

We then test the relative performance of the different approaches in a test set for fixed values: $\SNR\in\{20,40,60,80,100\}$ to better characterize (and discriminate) them, and for a fixed standard deviation of $\sigma=0.04$ corresponding to what was simulated in GREAT3 for the real galaxy space branch to obtain results on a representative observed galaxy set. The corresponding distribution of SNR in the last scenario is represented in Fig.~\ref{fig:histoSNR}. All the techniques are compared on exactly the same test sets.

For the ADMMnet approach when testing at different SNRs, we need to adjust the noise level in the galaxy images to the level of noise in the learning phase. We therefore rescale the galaxy images to reach this targeted noise level, based on noise level estimation in the images. This is performed via a robust standard procedure based on computing the median absolute deviation in the wavelet domain (using orthogonal daubechies wavelets with 3 vanishing moments).

\subsection{Quality criteria}
The performance of the deconvolution schemes is measured according to two different criteria, related to pixel error and shape measurement errors.
For pixel error we select a robust estimator:
\begin{equation}
\mathrm{P_{err}}\left(\widehat{\Xb}\right)=\mathrm{MED}\left(\frac{\|\widehat{\xb_\mathbf{i}}-\xb^{(t)}_\mathbf{i}\|^2_2}{\|\xb^{(t)}_\mathbf{i}\|_2^2}\right)_{i=1..n_g}
\end{equation}
where $\xb^{(t)}_\mathbf{i}$ is the targeted value, and with $\mathrm{MED}$ the median over the relative mean squared error computed for each galaxy $\xb_\mathbf{i}$  in the test set, in a central window of $41\times 41$ pixels common to all approaches.

For shape measurement errors, we compute the ellipticity using a KSB approach implemented in shapelens\footnote{\url{https://github.com/pmelchior/shapelens}} \citep{Kaiser1995,Viola2011}, that additionally computes an adapted circular weight function from the data.

We first apply this KSB method to the targets, taking as well into account the target isotropic gaussian PSF, to obtain reference complex ellipticities $\epsilon_i$ and windows. We then compute the complex ellipticity $\widehat{\epsilon_i}$ of the deconvolved galaxies using the same circular weight functions as their target counterpart. Finally, we compute
\begin{equation}
\mathrm{\epsilon_{err}}\left(\widehat{\Xb}\right)=\mathrm{MED}\left(\|\epsilon_i^{(t)} -\widehat{\epsilon_i}\|_2\right)_{i=1..n_g}
\end{equation}
to obtain a robust estimate of the ellipticity error in the windows set up by the target images, again in a central window of $41\times 41$ pixels common to all approaches..

We also report the distribution of pixel and ellipticity errors prior to applying the median when finer assessments need to be made. 
 
\section{Results}
\label{sec:Results}

\subsection{Setting the Tikhonet architecture and hyperparameters}

\begin{figure}[ht]\centering
  \includegraphics[width=\scalefig{0.9}]{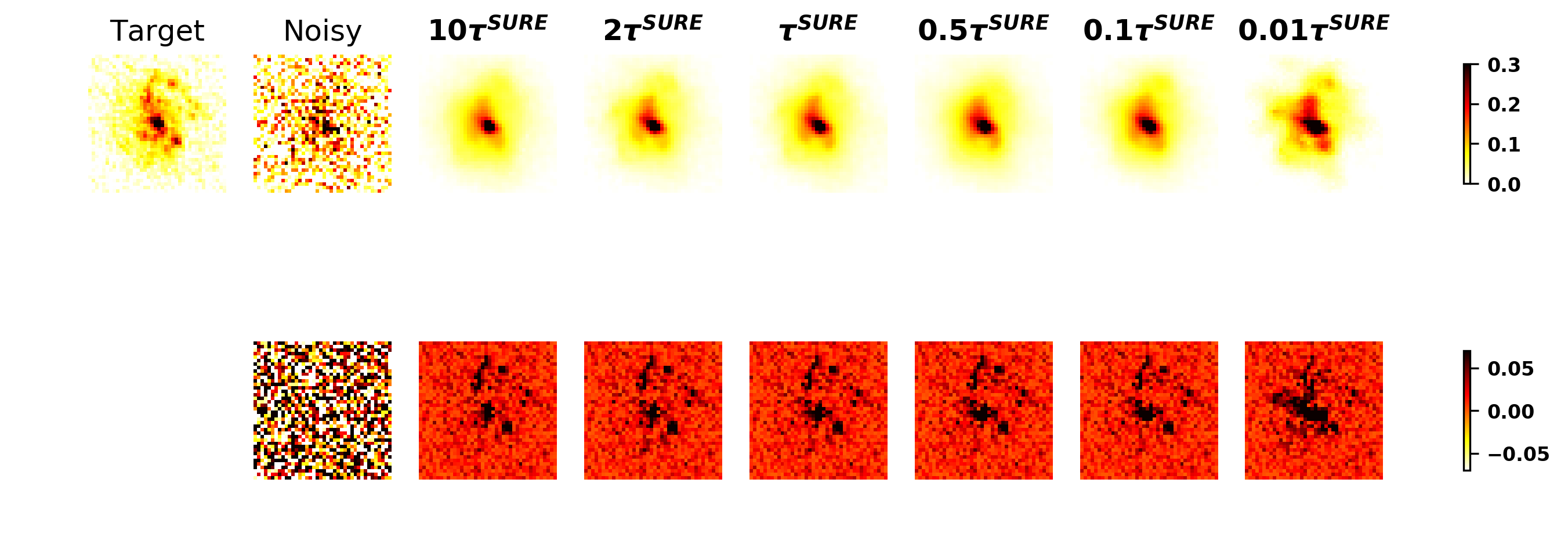}
  \caption{Visual impact of the hyperparameter choice for the Tikhonet approach at SNR20. Top: target and observations, followed by SURE estimates with different multiplicative factor. Bottom: residuals associated to the top row.}
  \label{fig:TikhoHyperVisu}
\end{figure}

For the Tikhonet, the key parameters to set are the hyperparameters $\lambda_i$ in Eq.~\ref{eq:SVTikhonovSolution}. In Fig.~\ref{fig:TikhoHyperVisu}, these hyperparameters are set to the parameters minimizing the SURE multiplied by factors ranging from 10 to 0.01 at $\SNR=20$, for the proposed X-Dense architecture (similar visual results are obtained for the "classical" U-Net). It appears that for the lowest factor, corresponding to the smallest regularization of deconvolution (i.e. more noise added in the deconvolved image), the Tikhonet is not able to perform as well as for intermediate values, in particular for exactly the SURE minimizer.

\begin{figure}[ht]\centering
  \includegraphics[width=\scalefig{0.47}]{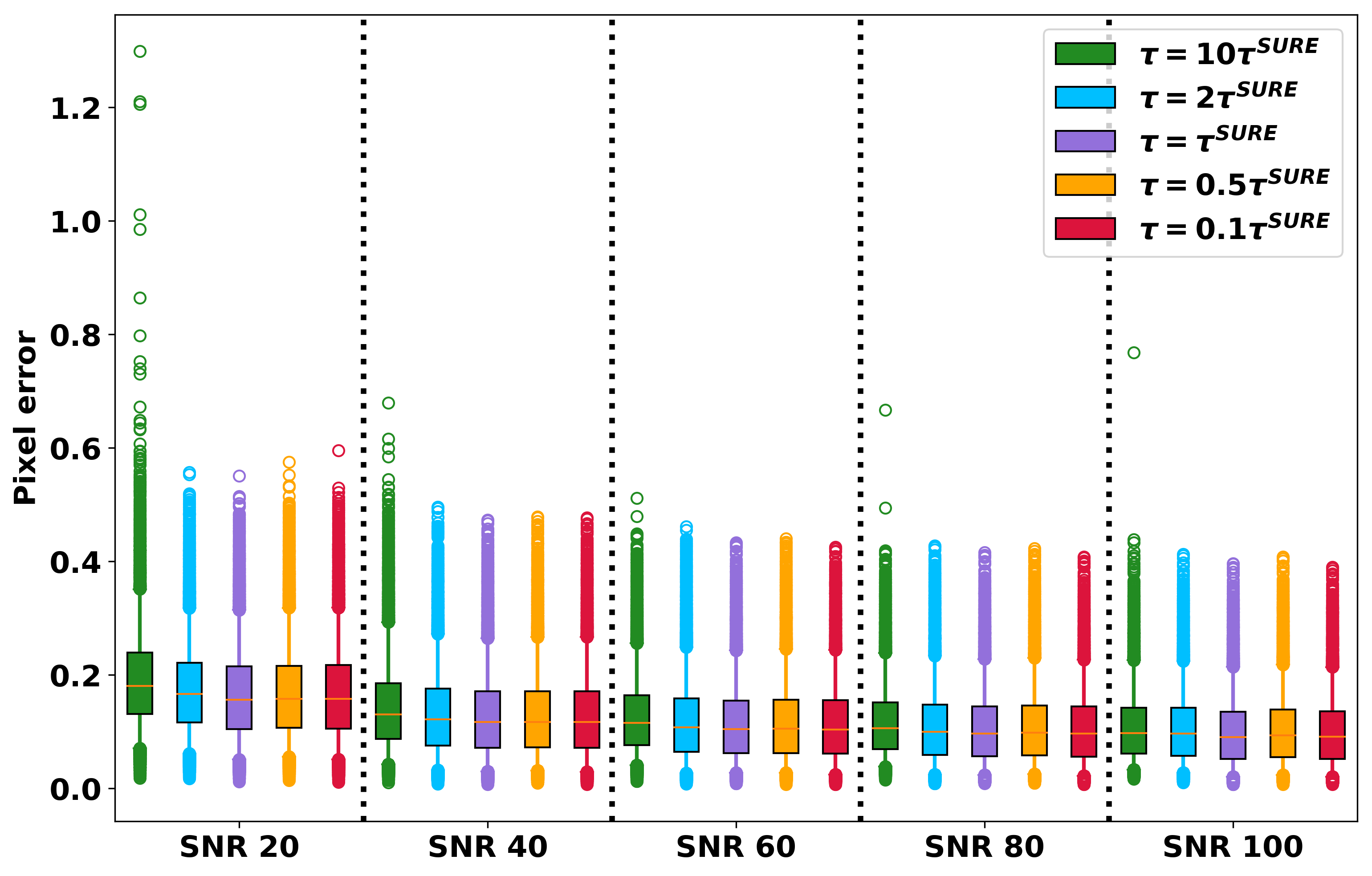}
  \includegraphics[width=\scalefig{0.47}]{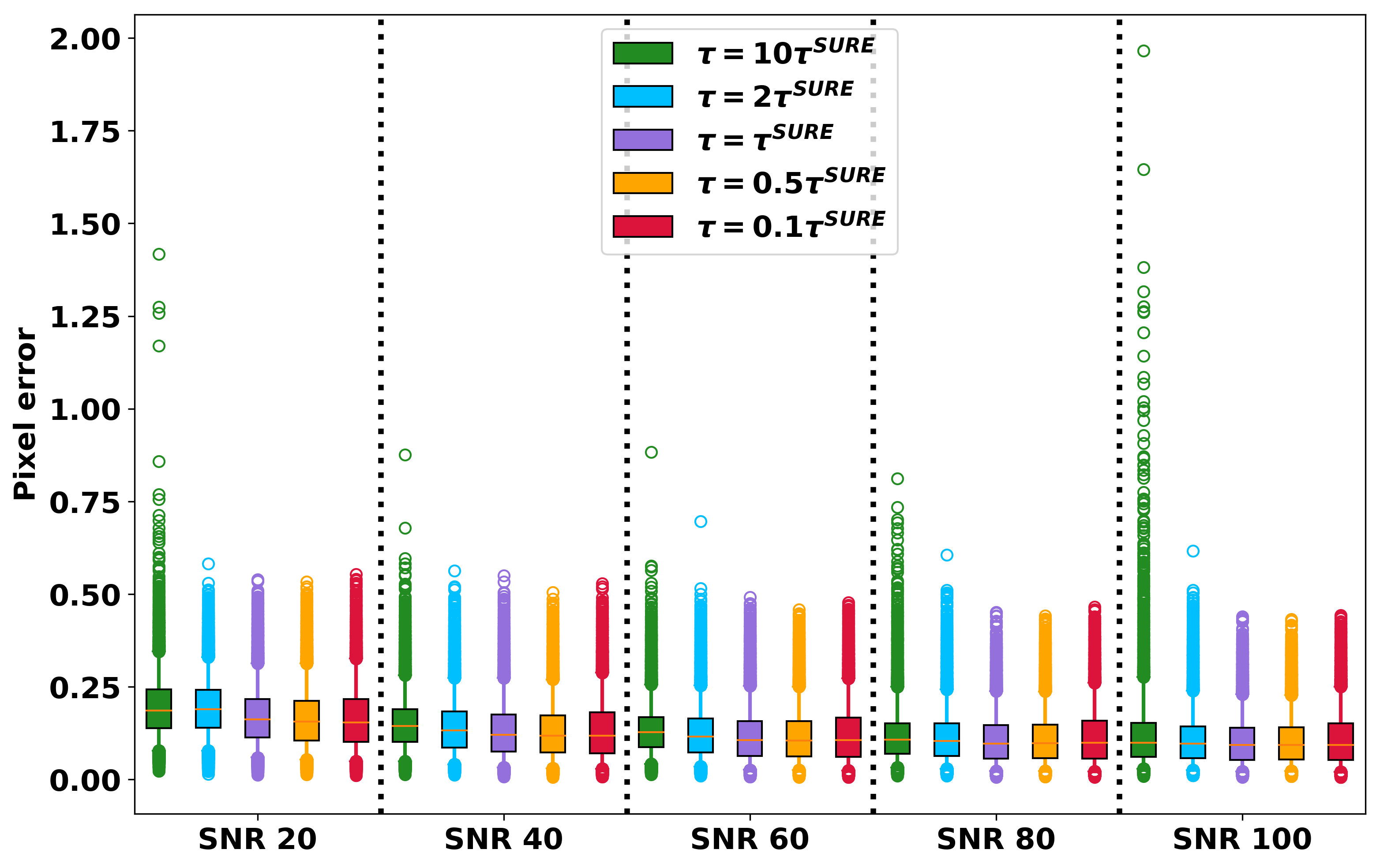}
  \caption{Impact of of the hyperparameter multiplicative factor value for the Tikhonet using the proposed XDense U-Net architecture (left) and "classical" U-Net architecture (right), in terms of pixel errors.The box indicate quartiles, while the vertical bars encompass $90\%$ of the data. Outliers are displayed with circles.}
  \label{fig:UnetXDenseTikhoHyperPixErr}
\end{figure}

\begin{figure}[ht]\centering
  \includegraphics[width=\scalefig{0.47}]{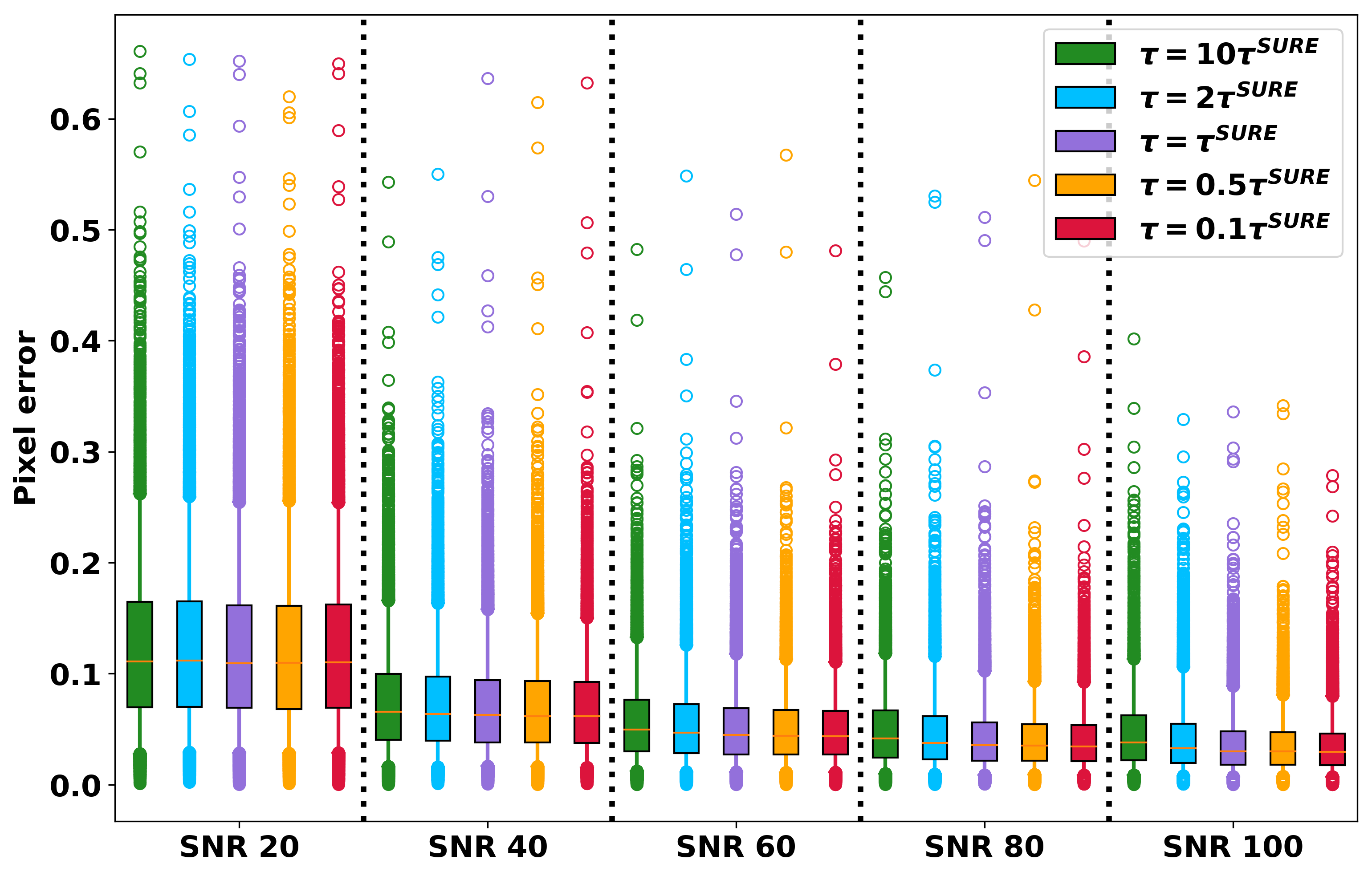}
  \includegraphics[width=\scalefig{0.47}]{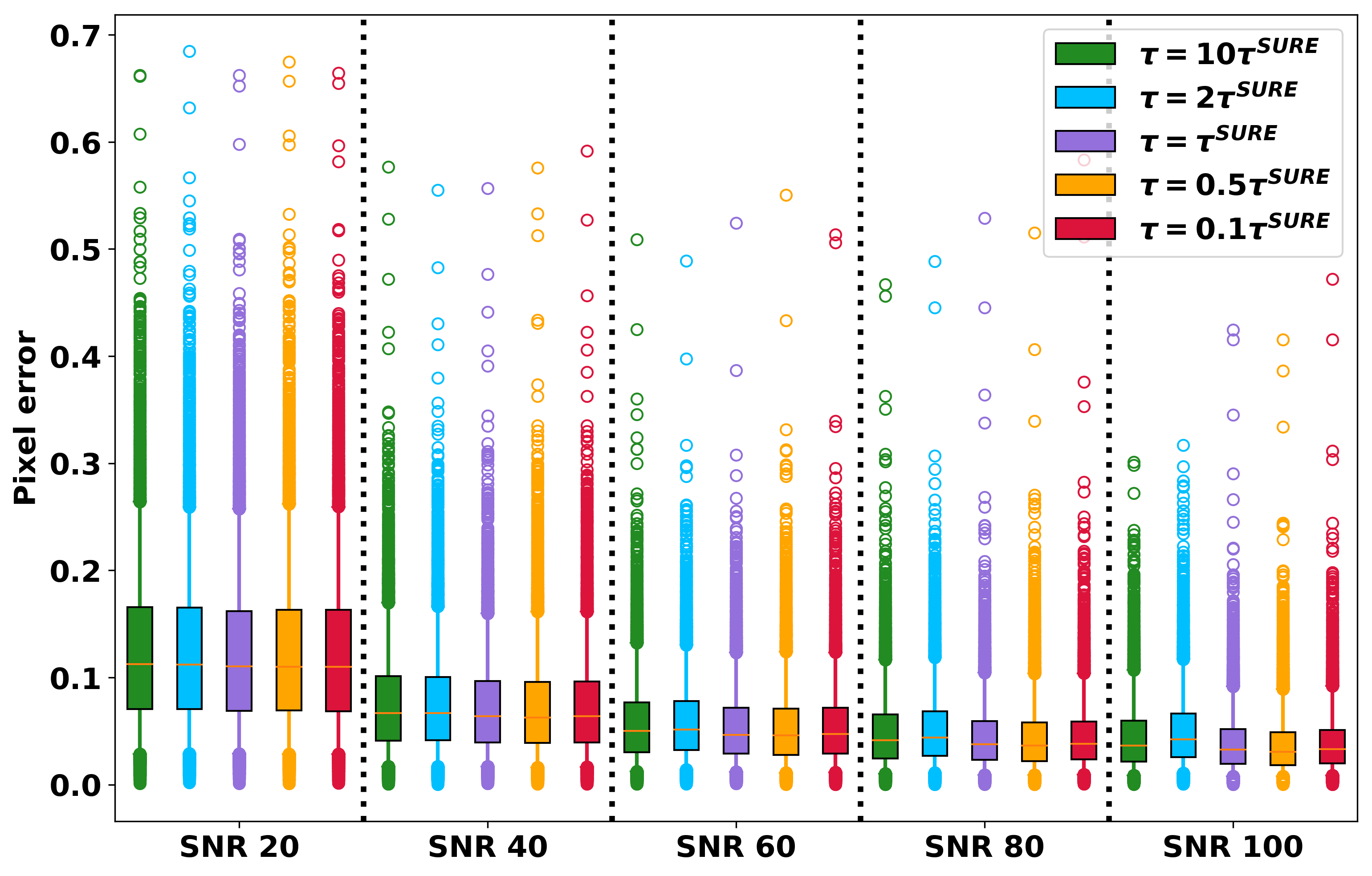}
  \caption{Impact of the hyperparameter multiplicative factor value for the Tikhonet using the proposed XDense U-Net architecture (left) and "classical" U-Net architecture (right), in terms of ellipticity errors.The box indicate quartiles, while the vertical bars encompass $90\%$ of the data. Outliers are displayed with circles.}
  \label{fig:UnetXDenseTikhoHyperEllErr}
\end{figure}

This is confirmed in Fig.~\ref{fig:UnetXDenseTikhoHyperPixErr} reporting the pixel errors for both proposed X-Dense and "classical" architecture, and Fig.~\ref{fig:UnetXDenseTikhoHyperEllErr} for the ellipticity errors. 

For both architectures, best results in terms of pixel or ellipticity errors are consistently obtained across all SNR tested for values of the multiplicative factor between 0.1 and 1. Higher multiplicative factors also lead to larger extreme errors in particular at low SNR. In the following, we therefore set this parameter to the SURE minimizer.
 
Concerning the choice of architecture, Fig.~\ref{fig:UnetXDenseTikhoHyperPixErr} illustrates that the XDense U-Net provides across SNR less extreme outliers in pixel errors for a multiplicative factor of 10, which is however far from providing the best results. Looking more closely at the median error values in Table~\ref{tbl:compUNetXDenseTikhonet} for the SURE minimizers, we see that slightly better results are consistently obtained for the proposed XDense U-Net architecture. In this experiment, the XDense obtains 4\% (resp. 3\%) less pixel errors  at $\SNR=20$ (resp. $\SNR=100$), and the most significant difference is an about 8\% improvement in ellipticity measurement at $\SNR=100$.

\begin{table}
\begin{center}
\caption{Comparison of U-Net architectures for the SURE selected hyperparameter. The first number is obtained with the XDense U-Net architecture, the second in parentheses with the "classical" U-Net architecture.}  
\begin{adjustbox}{max width=0.95\columnwidth}
{\renewcommand{\arraystretch}{1.5}
\begin{tabular}[c]{|l|l|l|l|l|l|} 
\hline
  & $\SNR=20$ & $\SNR=40$ & $\SNR=60$ & $\SNR=80$ & $\SNR=100$ \\[5pt]
\hline
\hline
Median Pixel Error  & \textbf{0.157} (0.163) & \textbf{0.117} (0.121) & \textbf{0.105} (0.106) &  \textbf{0.097} (0.097) & \textbf{0.090} (0.093) \\[5pt]
\hline
Median Ellipticity Error& \textbf{0.109} (0.110) & \textbf{0.063} (0.064) & \textbf{0.045} (0.046) &  \textbf{0.035} (0.038) & \textbf{0.030} (0.033) \\[5pt]
\hline 
\end{tabular}}
\end{adjustbox}
\label{tbl:compUNetXDenseTikhonet}
\end{center}
\end{table} 

\subsection{Setting the ADMMnet architecture and hyperparameters}

\begin{figure}[ht]\centering
  \includegraphics[width=\scalefig{0.9}]{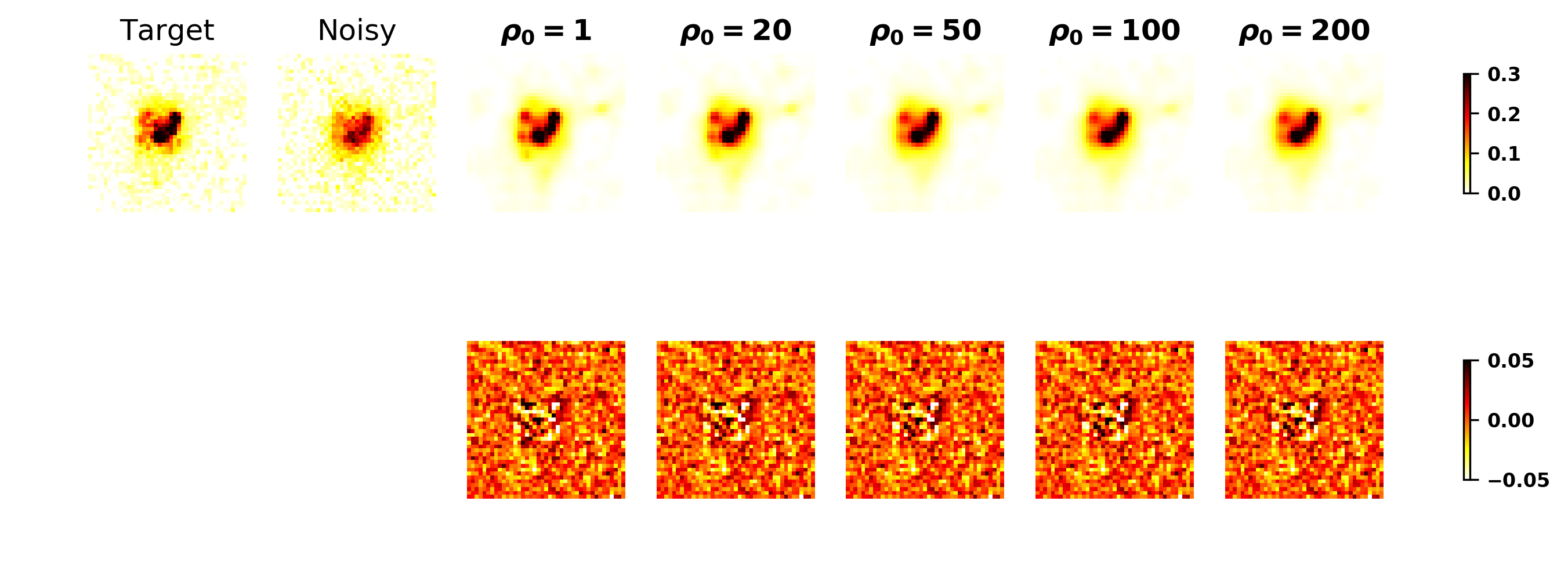}
  \caption{Visual impact of the initialization of $\rho$ for the ADMMnet for $\SNR=100$. Top: target and observations, followed by ADMM estimates with different augmented lagrangian parameter $\rho_0$. Bottom: residuals associated to the top row.}
  \label{fig:ADMMHyperVisu100}
\end{figure}

\begin{figure}[ht]\centering
  \includegraphics[width=\scalefig{0.9}]{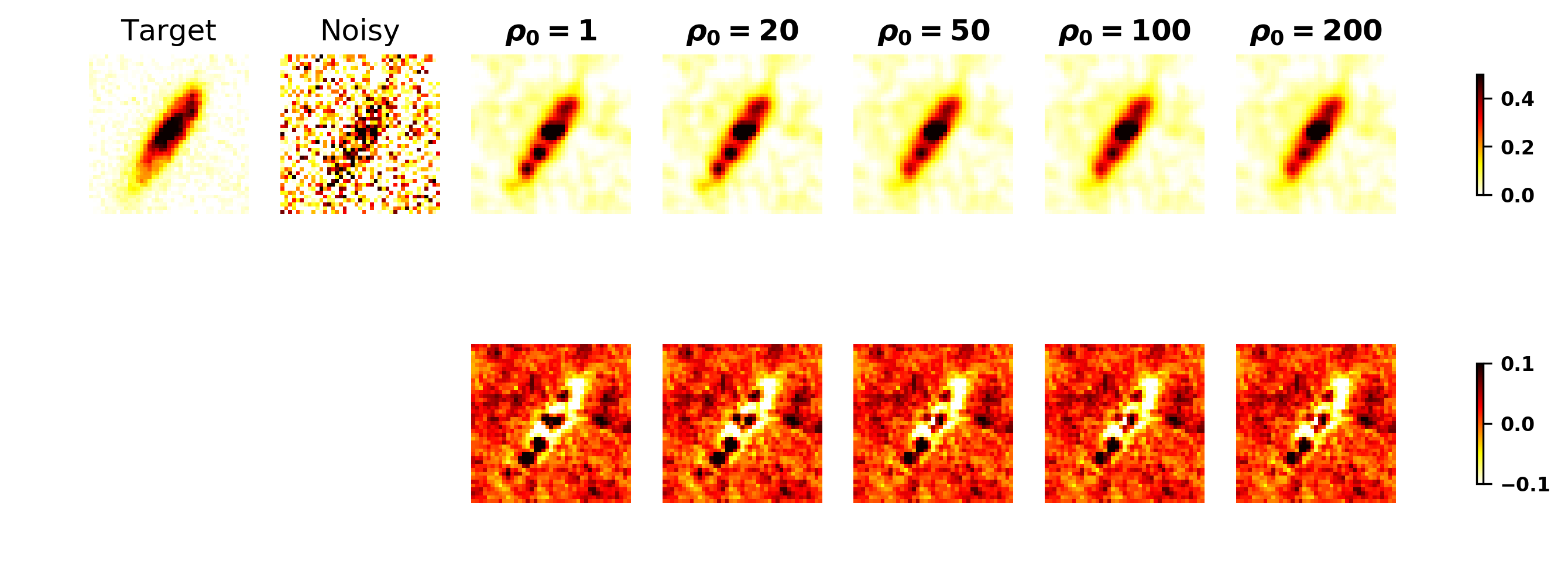}
  \caption{Visual impact of the initialization of $\rho$ for the ADMMnet for $\SNR=20$. Top: target and observations, followed by SURE estimates with different augmented lagrangian parameter $\rho_0$. Bottom: residuals associated to the top row.}
  \label{fig:ADMMHyperVisu20}
\end{figure}

For the ADMMnet, we set manually the hyperparameters $\rho_{max}=200$, $\epsilon=0.01$ to lead to ultimate stabilization of the algorithm, $\eta=0.5$ and $\gamma=1.4$ to explore intermediate $\rho$ values, and we investigate the choice of parameter $\rho_0$ to illustrate the impact of the continuation scheme on the solution. This is illustrated in Fig.~\ref{fig:ADMMHyperVisu100} at high SNR, and Fig.~\ref{fig:ADMMHyperVisu20} at low SNR for the proposed XDense U-Net architecture. When $\rho_0$ is small, higher frequencies are recovered in the solution as illustrated in galaxy substructures in Fig.~\ref{fig:ADMMHyperVisu100}, but this could lead to artefacts at low SNR as illustrated in Fig.~\ref{fig:ADMMHyperVisu20}.

\begin{figure}[ht]\centering
  \includegraphics[width=\scalefig{0.4}]{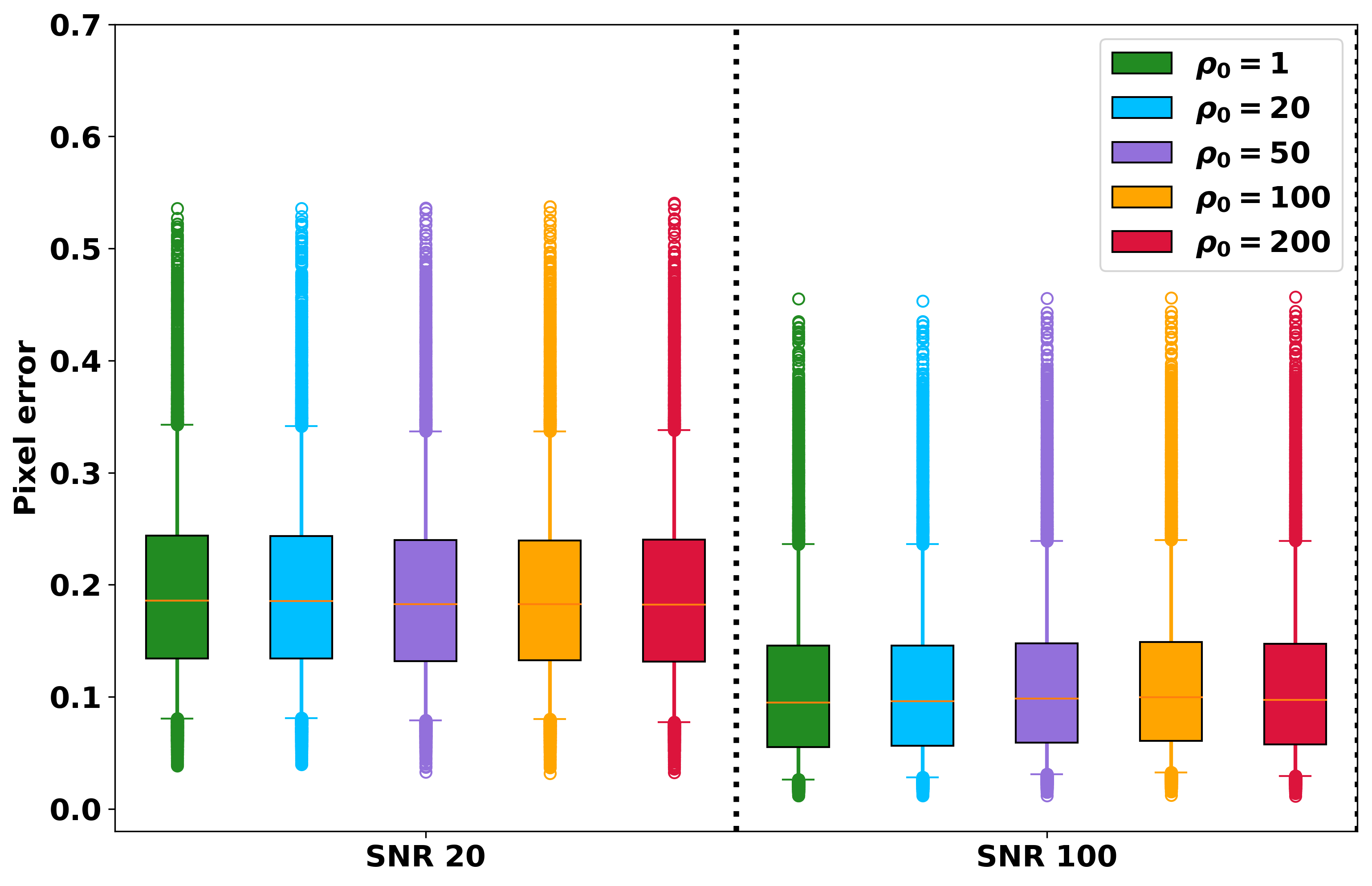}
  \includegraphics[width=\scalefig{0.4}]{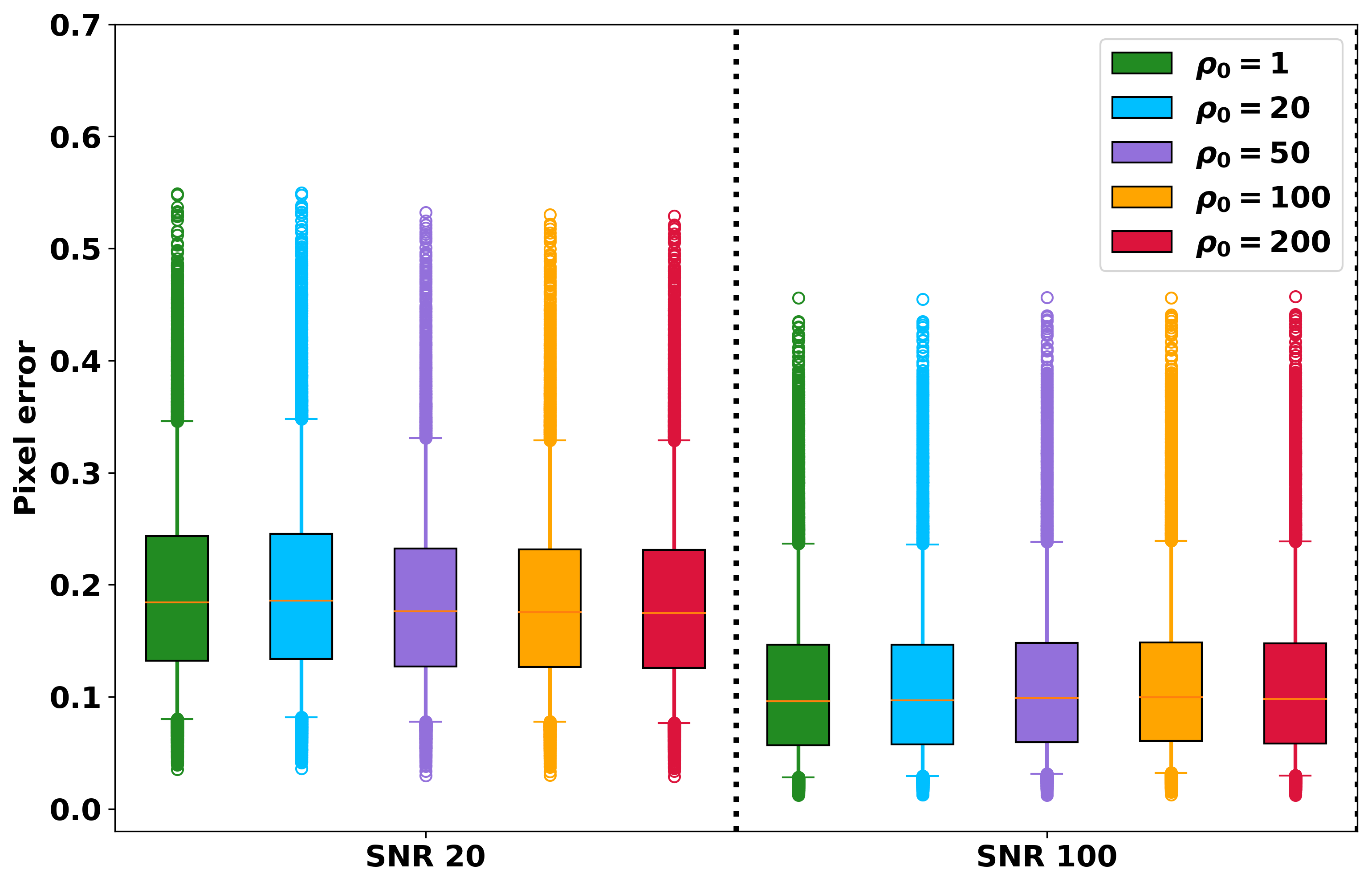}
  \caption{Impact of the hyperparameter $\rho_0$ value for the ADMMnet, in terms of pixel error, for the proposed XDense U-Net  (left) and "classical" U-Net (right).The box indicate quartiles, while the vertical bars encompass $90\%$ of the data. Outliers are displayed with circles.}
  \label{fig:ADMMHyperStatsPixErr}
\end{figure}

\begin{figure}[ht]\centering
  \includegraphics[width=\scalefig{0.4}]{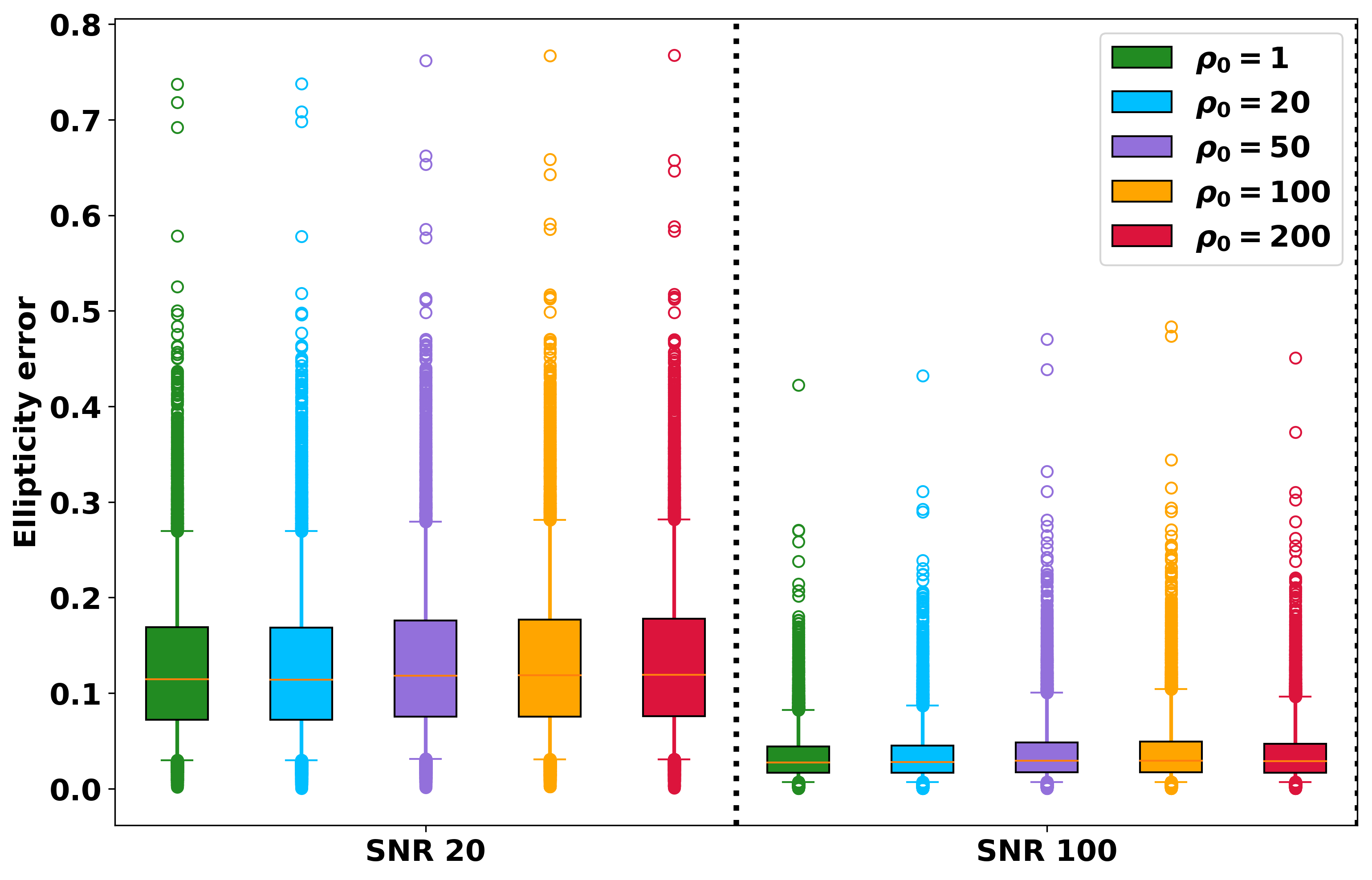}
  \includegraphics[width=\scalefig{0.4}]{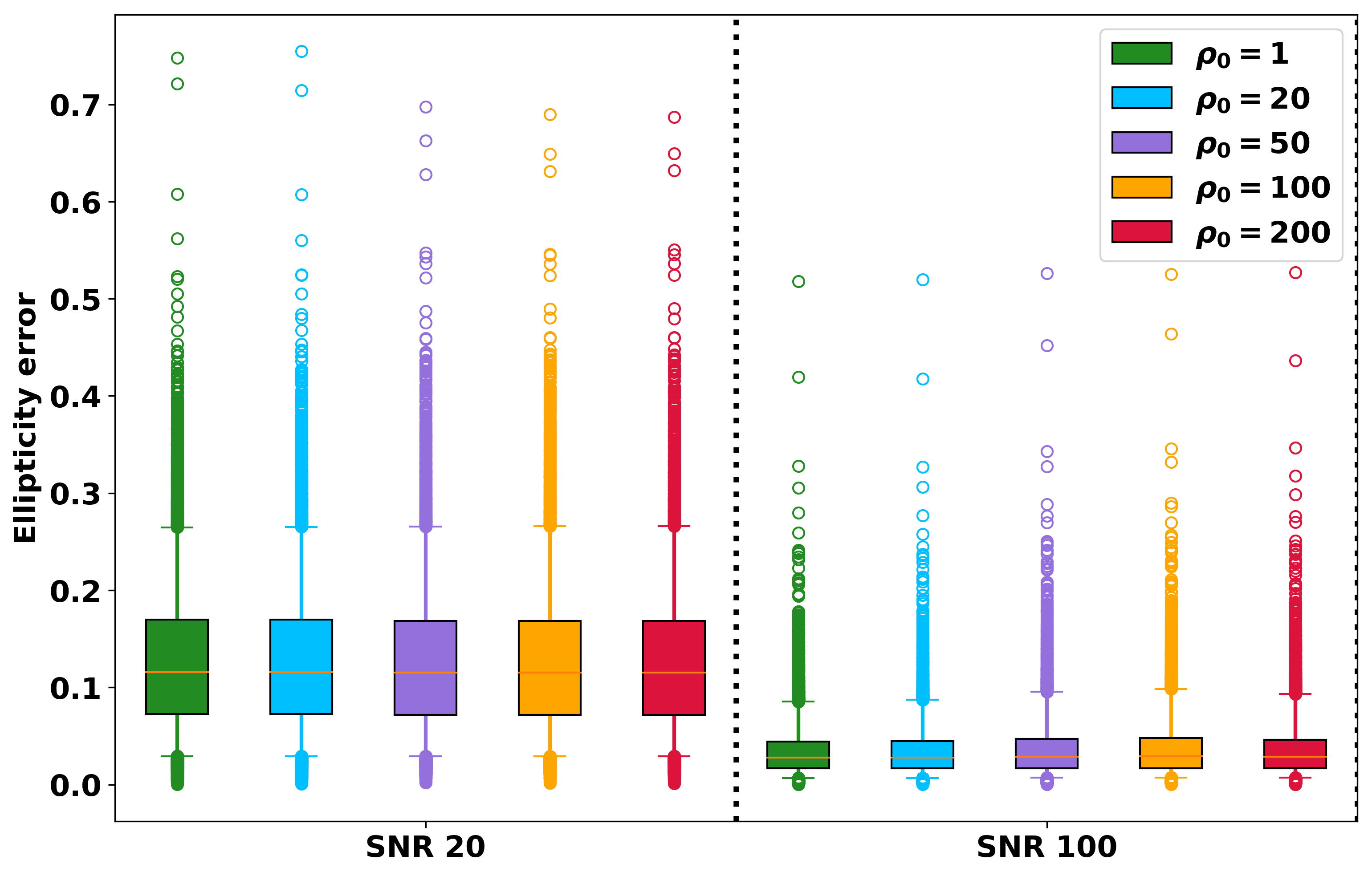}
  \caption{Impact of the hyperparameter choice $\rho_0$ for the ADMMnet, in terms of ellipticity error, for the proposed XDense- U-Net  (left) and "classical" U-Net (right).The box indicate quartiles, while the vertical bars encompass $90\%$ of the data. Outliers are displayed with circles.}
  \label{fig:ADMMHyperStatsEllErr}
\end{figure}

Quantitative results concerning the two architectures are presented in Fig.~\ref{fig:ADMMHyperStatsPixErr} for pixel errors and Fig.~\ref{fig:ADMMHyperStatsEllErr} for ellipticity errors. The distribution of errors is very stable with respect to the hyperparameter $\rho_0$ value, and similar for both architectures.
 
When looking at pixel error at high SNR and low SNR for both architecture, the lowest pixel errors in terms of median are obtained at low SNR for larger $\rho_0$, while at high SNR $\rho_0=1$ is the best. In terms of ellipticity errors,  $\rho_0=1$ allows to obtain consistently the best results at low and high SNR.

To better compare the differences between the two architectures, the median errors are reported in Table~\ref{tbl:compUNetXDenseADMMNet}. At low SNR the "classical" U-Net performance varies more than that of the XDense and at $\SNR=20$, best results are obtained for the "classical" U-Net approach ($4\%$ improvement over X-Dense). At high SNR however, best results are consistently obtained across $\rho_0$ values with the proposed XDense U-Net (but only $1\%$ improvement over the U-Net for the best $\rho_0=1$). Finally, concerning ellipticity median errors, best results are obtained for the smallest value $\rho_0=1$ for both architectures and the proposed XDense U-Net performs slightly better than the "classical" U-Net (about $1\%$ better both at low and high SNR).
 
 \begin{table}
\begin{center}
\caption{Comparison of U-Net architectures for median errors. The first number is obtained with the proposed XDense U-Net architecture, the second in parentheses with the "classical" U-Net architecture.} 
\begin{adjustbox}{max width=0.95\columnwidth}
{\renewcommand{\arraystretch}{1.5}
\begin{tabular}[c]{|l|l|l|l|l|l|l|l|l|l|l|} 
\hline
  & \multicolumn{5}{|c|}{$\SNR=20$} \\[5pt]
\hline
  & $\rho_0=1$ & $\rho_0=20$ & $\rho_0=50$ & $\rho_0=100$& $\rho_0=200$ \\[5pt]
\hline
Median Pixel Error  & 0.186 (\textbf{0.184}) & \textbf{0.185} (0.186) & 0.182 (\textbf{0.176}) &  0.183 (\textbf{0.175}) & 0.182 (\textbf{0.175})  \\[5pt]
\hline
Median Ellipticity Error& \textbf{0.114} (0.116) & \textbf{0.114} (0.116) & 0.118 (\textbf{0.115}) & 0.119 (\textbf{0.115}) & 0.119 (\textbf{0.115}) \\[5pt]
\hline
\hline
& \multicolumn{5}{|c|}{$\SNR=100$}\\[5pt]
\hline
 & $\rho_0=1$ & $\rho_0=20$ & $\rho_0=50$ & $\rho_0=100$ & $\rho_0=200$\\[5pt]
\hline
Median Pixel Error  & \textbf{0.095} (0.096) & \textbf{0.096} (0.097) & \textbf{0.098} (0.099) &  \textbf{0.099} (0.099) & \textbf{0.097} (0.098) \\[5pt]
\hline
Median Ellipticity Error & \textbf{0.028} (0.028) & 0.028 (\textbf{0.028}) & 0.029 (\textbf{0.029}) &  0.029 (\textbf{0.029}) & 0.029 (\textbf{0.028}) \\[5pt]

\hline 
\end{tabular}}
\end{adjustbox}
\label{tbl:compUNetXDenseADMMNet}
\end{center}
\end{table} 

Overall, this illustrates that the continuation scheme has a small impact in particular on the ellipticity errors, and that best results are obtained for different $\rho_0$ and network architectures if pixel or ellipticity errors are considered, depending on the SNR. The "classical" U-Net allows smaller pixel errors than the proposed XDense at low SNR, but also leads to slightly higher pixel errors at higher SNR and ellipticity errors at both low and high SNR.
In practice we keep in the following for the proposed XDense U-Net approach $\rho_0=1$ for further comparison with other deconvolution approaches as the pixel error is varying slowly with this architecture as a function of $\rho_0$.
 
\subsection{DNN versus sparsity and low-rank}
 
We compare our two deep learning schemes with the XDense U-Net architecture and the hyperparameters set as described in the previous sections with the sparse and the low rank approaches of \citet{Farrens2017}, implemented in sf\_deconvolve \footnote{\url{https://github.com/sfarrens/sf_deconvolve}}. For the two methods, we used all parameters selected by default, reconvolved the recovered galaxy images with the target PSF and selected the central $41\times 41$ pixels of the observed galaxies to be processed in particular to speed up the computation of the singular value decomposition used in the low rank constraint (and therefore of the whole algorithm) as in \citet{Farrens2017}. All comparisons are made in this central region of the galaxy images.

We now illustrate the results for a variety of galaxies recovered at different SNR for the sparse, low-rank deconvolution approaches and the Tikhonet and ADMMnet. 

\begin{figure}[ht]\centering
  \includegraphics[width=\scalefig{0.95}]{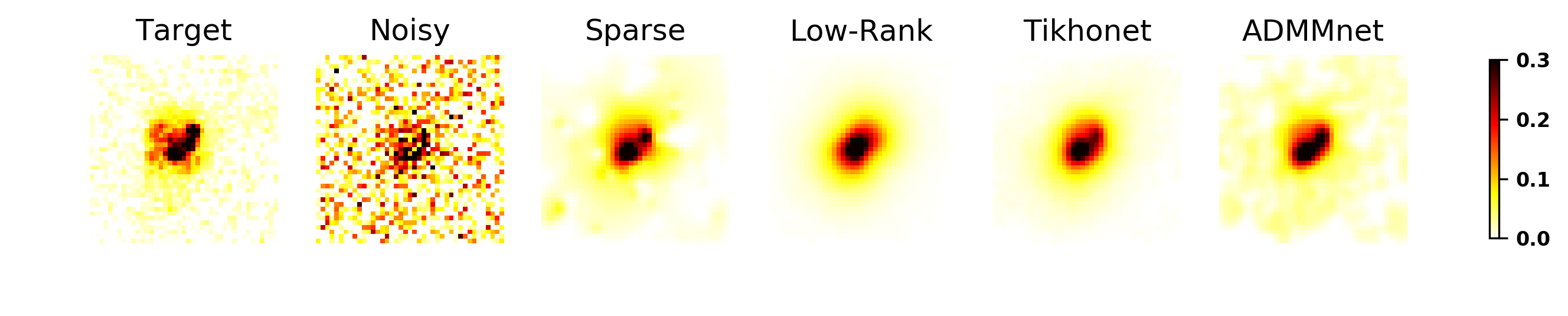}
  \includegraphics[width=\scalefig{0.95},trim={0 0 0 0.7cm},clip]{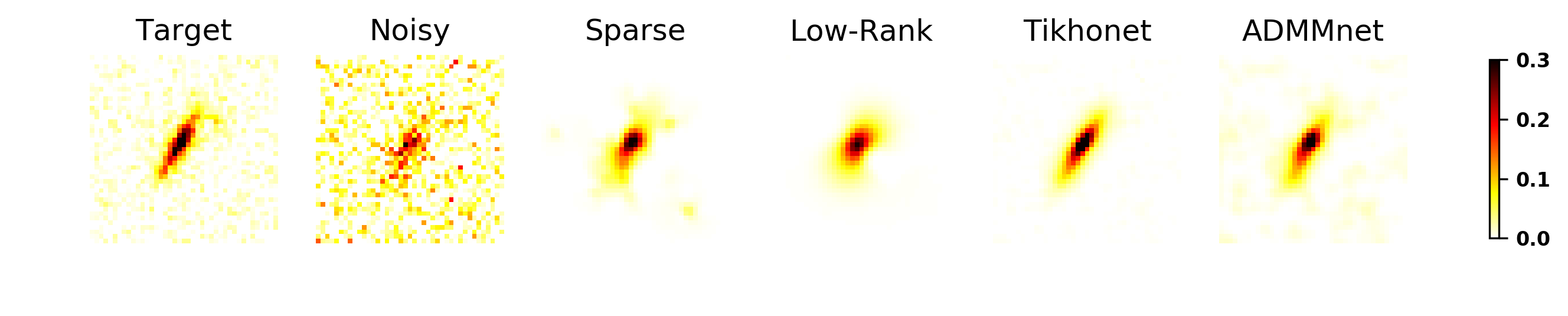}
  \includegraphics[width=\scalefig{0.95},trim={0 0 0 0.7cm},clip]{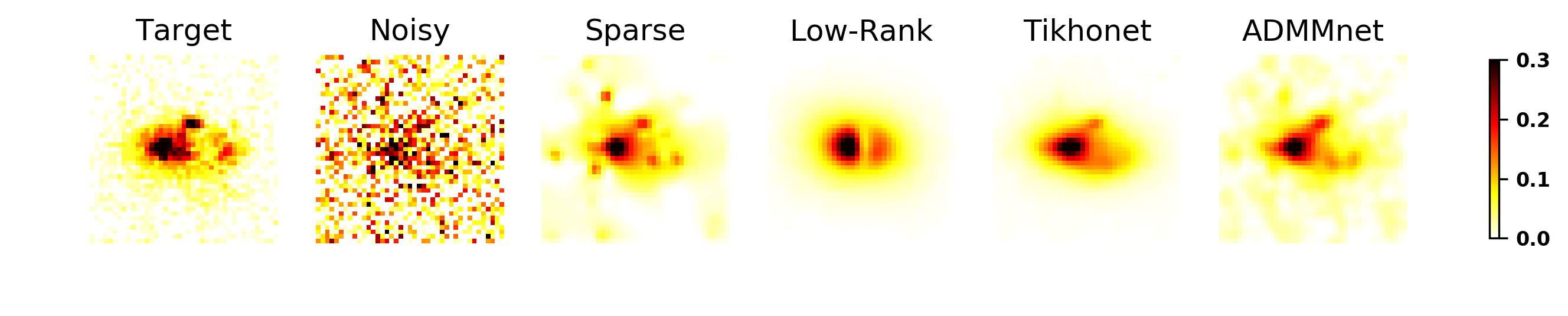}
  \includegraphics[width=\scalefig{0.95},trim={0 0 0 0.7cm},clip]{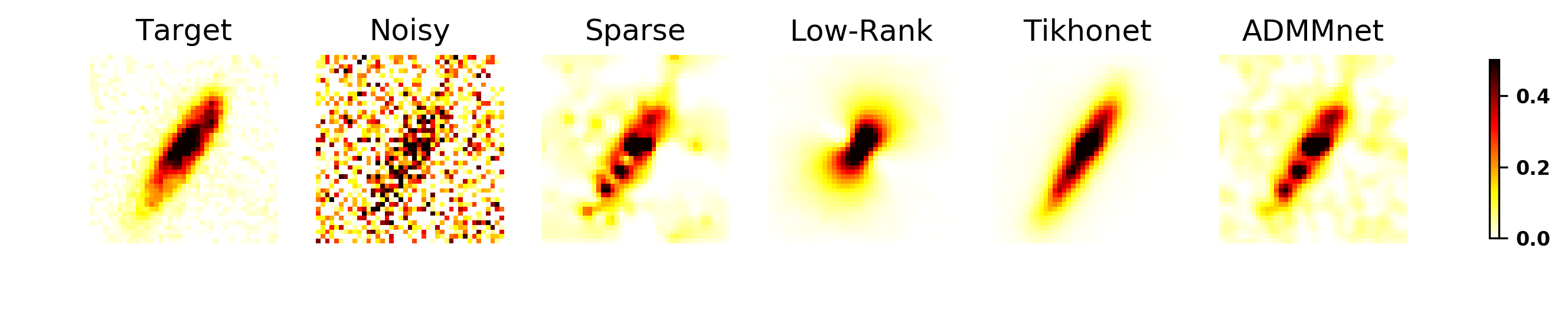}
  \includegraphics[width=\scalefig{0.95},trim={0 0 0 0.7cm},clip]{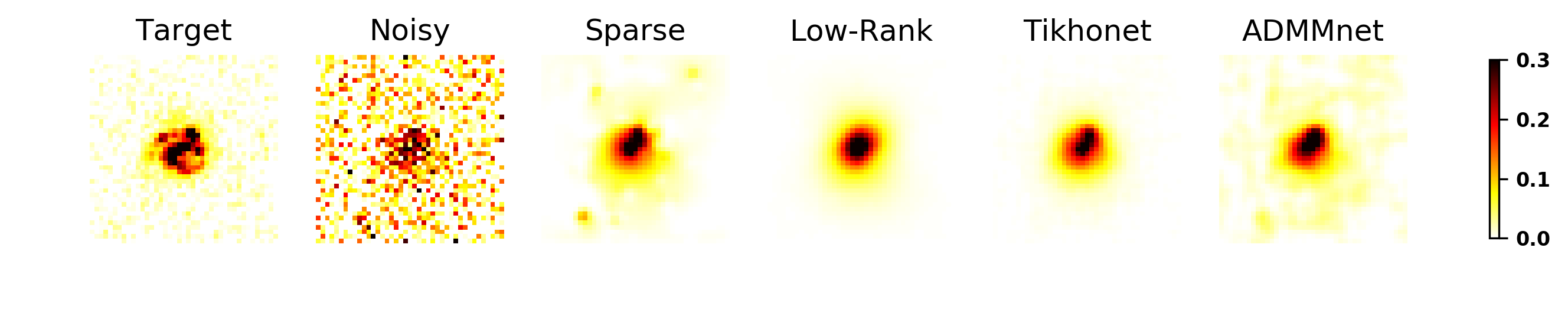}
  \includegraphics[width=\scalefig{0.95},trim={0 0 0 0.7cm},clip]{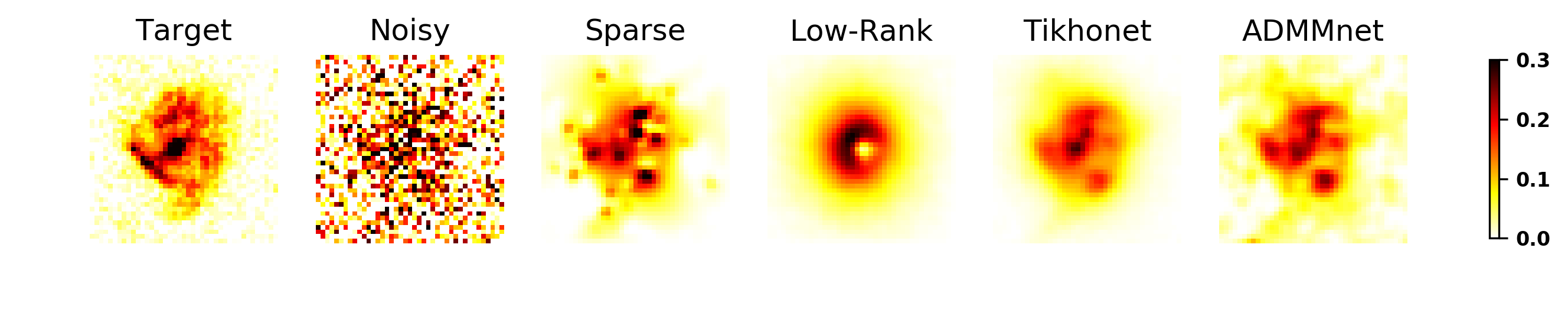}
  \caption{Deconvolved images with the various approaches for $\SNR=20$. Each row corresponds to a different processed galaxy. From left to right: image to recover, observation with a noise realization, sparse and low rank approaches, and finally Tikhonet and ADMMnet results.}
  \label{fig:galimSNR20}
\end{figure}

We first display several results at low SNR ($\SNR=20$) in Fig.~\ref{fig:galimSNR20} to illustrate the robustness of the various deconvolution approaches. Important artefacts appear in the sparse approach, illustrating the difficulty of recovering the galaxy images in this high noise scenario: retained noise in the deconvolved images lead to these point-like artefacts.

For the low rank approach, low frequencies seems to be partially well recovered, but artefacts appears for elongated galaxies in the direction of the minor axis. Finally, both Tikhonet and ADMMnet seem to recover better the low frequency information, but the galaxy substructures are essentially lost. The ADMMnet seems to recover in this situation sharper images but with more propagated noise/artefacts than the Tikhonet, with similar features as for the sparse approach but with less point-like artefacts.

\begin{figure}[ht]\centering
  \includegraphics[width=\scalefig{0.95}]{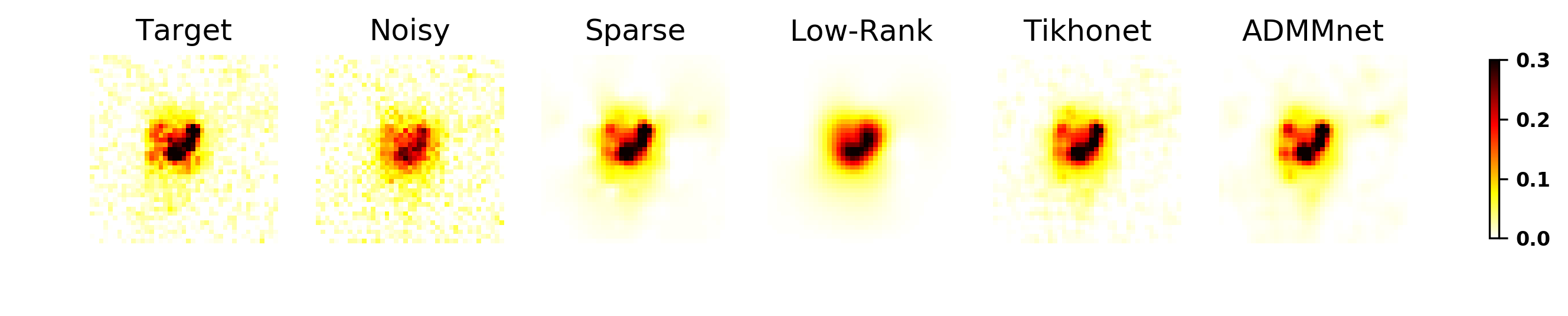}
  \includegraphics[width=\scalefig{0.95},trim={0 0 0 0.7cm},clip]{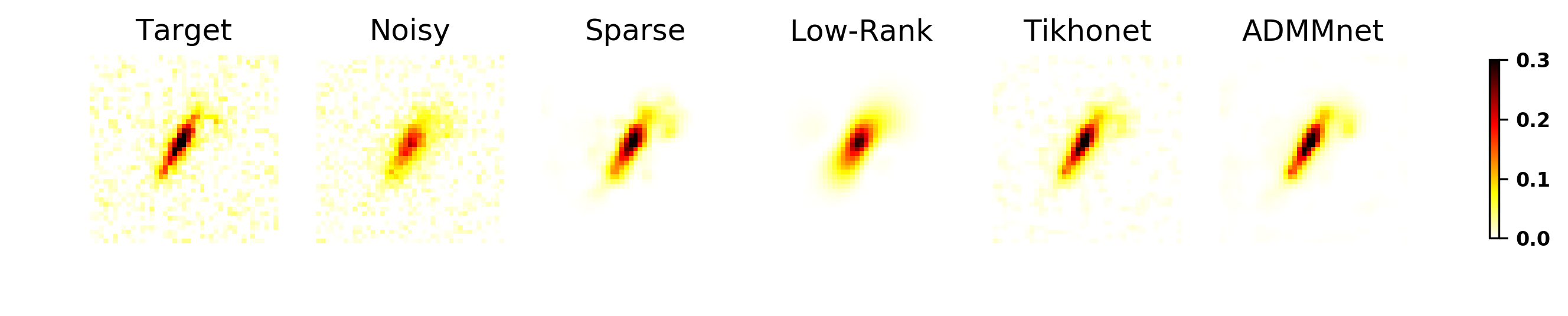}
  \includegraphics[width=\scalefig{0.95},trim={0 0 0 0.7cm},clip]{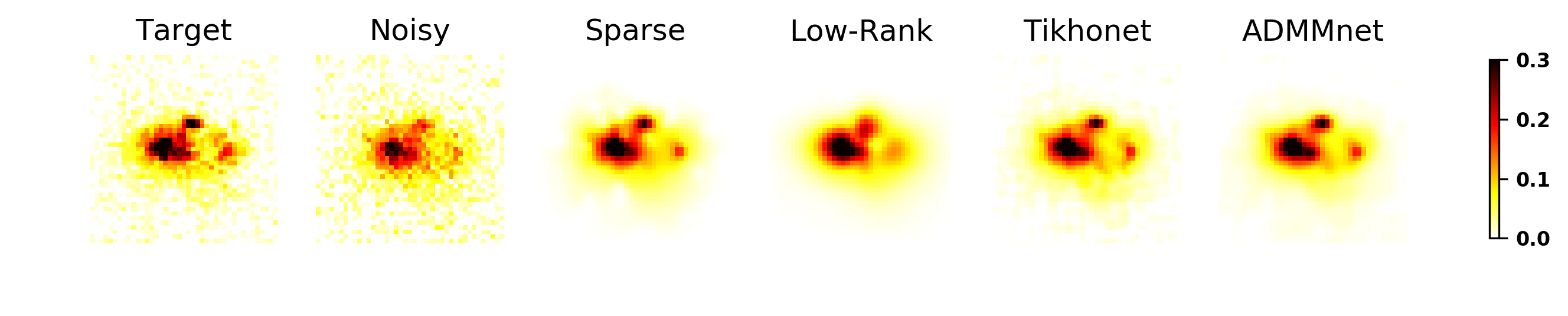}
  \includegraphics[width=\scalefig{0.95},trim={0 0 0 0.7cm},clip]{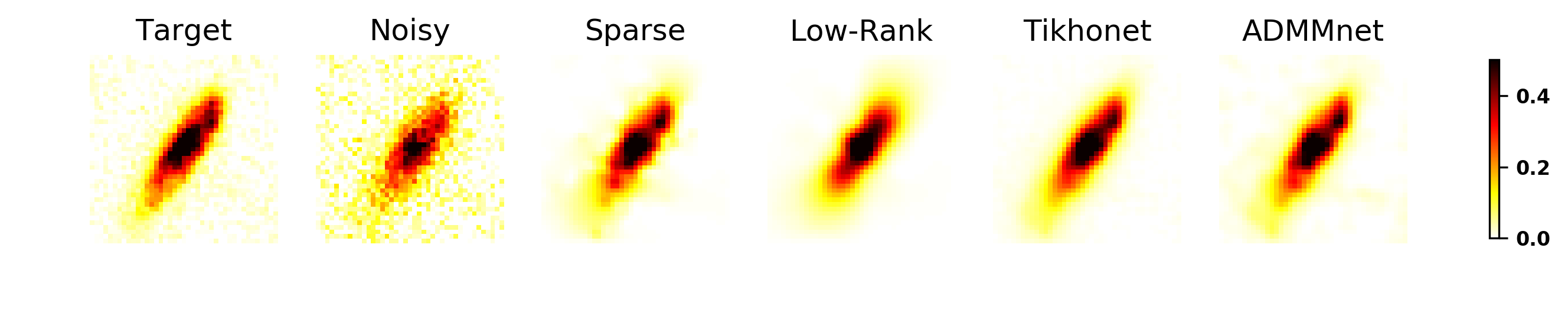}
  \includegraphics[width=\scalefig{0.95},trim={0 0 0 0.7cm},clip]{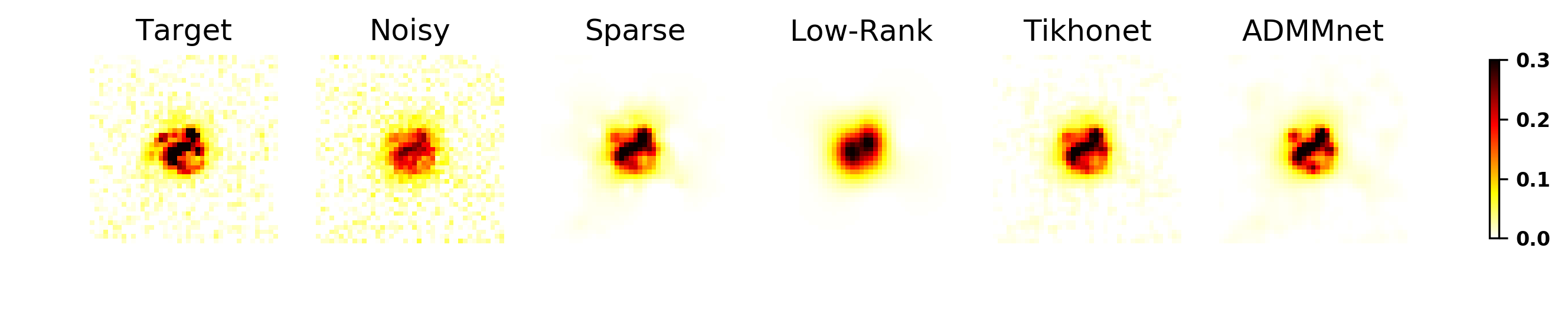}
  \includegraphics[width=\scalefig{0.95},trim={0 0 0 0.7cm},clip]{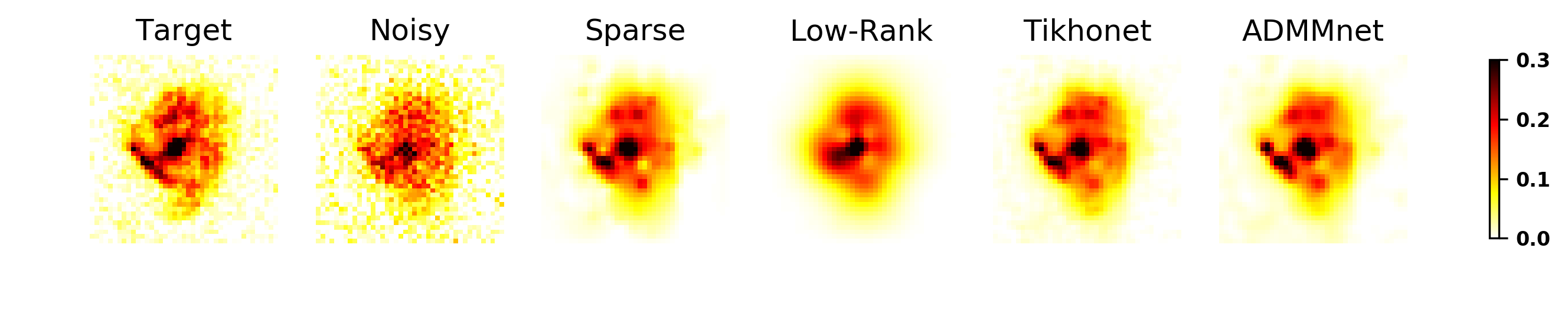}
  \caption{Deconvolved images with the various approaches for $\SNR=100$. Each row corresponds to a different processed galaxy. From left to right: image to recover, observation with a noise realization, sparse and low rank approaches, and finally Tikhonet and ADMMnet results.}
  \label{fig:galimSNR100}
\end{figure}

We also display these galaxies at a much higher SNR ($\SNR=100$) in Fig.~\ref{fig:galimSNR100} to assess the ability of the various deconvolution schemes to recover galaxy substructures in a low noise scenario.

The low-rank approach displays less artefacts than at low SNR, but still does not seem  to be able to adequately represent elongated galaxies or directional substructures in the galaxy. This is probably due to the fact that the low rankness approach does not adequately cope with translations, leading to over-smooth solutions. On the contrary,  Tikhonet, ADMMnet and sparse recovery lead to recover substructures of the galaxies. 

Overall the two proposed deconvolution approaches using DNNs lead to the best visual results across SNR.

\begin{figure}[ht]\centering
  \includegraphics[width=\scalefig{0.45}]{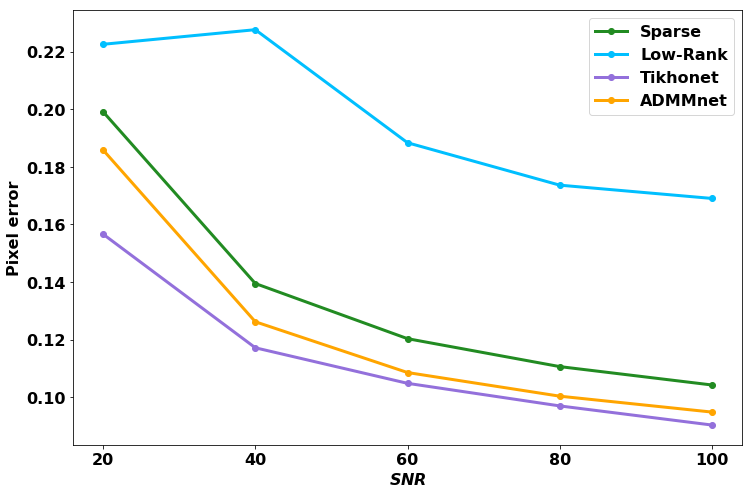}
  \includegraphics[width=\scalefig{0.45}]{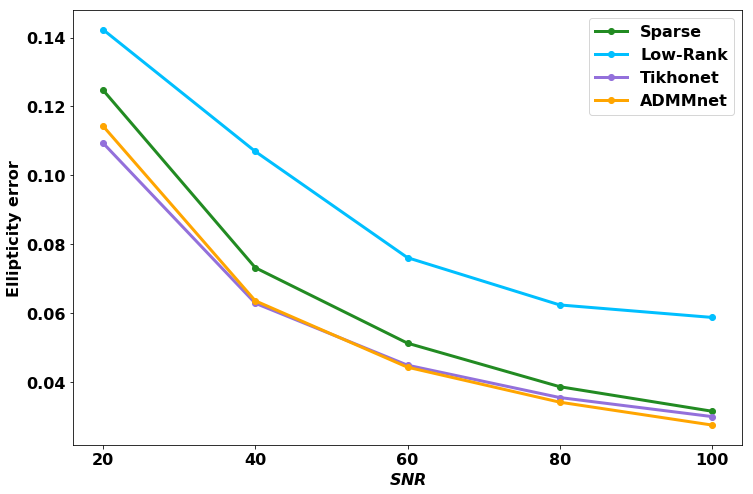}
  \caption{Deconvolution quality criteria for the different deconvolution schemes. Left: median pixel error, Right: median ellipticity error.}
  \label{fig:statAllMethods}
\end{figure}

The quantitative deconvolution criteria are presented in Fig.~\ref{fig:statAllMethods}. Concerning median pixel error, this figure illustrates that both Tikhonet and ADMMnet perform better than the sparse and low-rank approach to recover the galaxy intensity values, whatever the SNR. In these noise settings the low-rank approach performed consistently worst than using sparsity. In terms of pixel errors, the sparse approach median errors are $27\%$ (resp. $15\%$) larger at $\SNR=20$ (resp. $\SNR=100$) compared to the Tikhonet results.  The Tikhonet seems to perform slightly better than the ADMMnet with this criterion as well. 

For shape measurement errors, the best results are obtained with the Tikhonet approach at low SNR (up to $\SNR=40$), and then the ADMMnet outperforms the others at higher SNR. In terms of ellipticity errors, the sparse approach median errors are $14\%$ (resp. $5\%$) larger at $\SNR=20$ (resp. $\SNR=100$) compared to the Tikhonet results. Finally the low-rank performs the worst whatever the SNR. To summarize, these results clearly favour the choice of the DNN approaches resulting in lower errors consistently across SNR.

This is confirmed when looking at a realistic distribution of galaxy SNR, as shown in Table~\ref{tbl:cstNoise}. In terms of both median pixel and ellipticity errors, the proposed deep learning approaches perform similarly, and outperforms both sparse and low-rank approaches: median pixel error is reduced by almost $14\%$ (resp. $9\%$) for the Tikhonet (resp. ADMMnet) approach compared to sparse recovery, and ellipticity errors by about $13\%$ for both approaches. Higher differences are observed for the low-rank approach.

\begin{table}
\begin{center}
\caption{Criteria for constant noise simulations ($\sigma=0.04$). Best results are indicated in bold.} 
\begin{adjustbox}{max width=0.95\columnwidth}
{\renewcommand{\arraystretch}{1.5}
\begin{tabular}[c]{|l|l|l|l|l|} 
\hline
  & Sparse & Low-Rank & Tikhonet & ADMMnet \\[5pt]
\hline
\hline
Median Pixel Error  & 0.130 & 0.169 &\textbf{0.112} &  0.119 \\[5pt]
\hline
Median Ellipticity Error& 0.061 & 0.072&\textbf{0.053}& 0.054 \\[5pt]
\hline 
\end{tabular}}
\end{adjustbox}
\label{tbl:cstNoise}
\end{center}
\end{table}

\subsection{Computing Time}

Finally, we also report in Table~\ref{tbl:timing} the time necessary to learn the networks and process the set of $10000$ galaxies on the same GPU/CPUs, as this is a crucial aspect when potentially processing a large number of galaxies such as in modern surveys. Among DNNs, learning the parameters of the denoising network for the ADMMnet is faster than those of the post-processing network in the Tikhonet since the latter requires each batch to be deconvolved. However once the network parameters have been learnt, the Tikhonet based on a closed-form deconvolution is the fastest to process a large number of galaxies (about 0.05s per galaxy). On the other hand, learning and restoring 10000 galaxies is quite fast for the low-rank approach, while iterative algorithms such as ADMMnet or the primal-dual algorithm for sparse recovery are similar in terms of computing time (about 7 to 10s per galaxy). All these computing times could however be reduced if the restoration of different galaxy images is performed in parallel, which has not been implemented.

\begin{table}
\begin{center}       
\caption{Computing time for the various approaches (in hours).} 
\begin{tabular}{|l|l|l|} 
\hline
\rule[-1ex]{0pt}{3.5ex}    Method & Learning & Processing 10000 galaxies  \\
\hline
\hline
\rule[-1ex]{0pt}{3.5ex}  Sparse  & / & 24.7  \\
\hline
\rule[-1ex]{0pt}{3.5ex}  Low-rank& \multicolumn{2}{|c|}{5.2} \\
\hline
\rule[-1ex]{0pt}{3.5ex}  Tikhonet& 21.5 & 0.1\\
\hline
\rule[-1ex]{0pt}{3.5ex}  ADMMnet& 16.2 & 20.3 \\
\hline 
\end{tabular}
\label{tbl:timing}
\end{center}
\end{table} 


\section{Conclusions}
\label{sec:ccl}

We have proposed two new space-variant deconvolution strategies for galaxy images based on deep neural networks, while keeping all knowledge of the PSF in the forward model: the Tikhonet, a post-processing approach of a simple Tikhonov deconvolution with a DNN, and the ADMMnet based on regularization by a DNN denoiser inside an iterative ADMM PnP algorithm for deconvolution. We proposed to use for galaxy processing a DNN architecture based on the U-Net particularly adapted to deconvolution problems, with small modifications implemented (dense Blocks of separable convolutions, and no skip connection) to lower the number of parameters to learn compared to a "classical" U-Net implementation. We finally evaluated these approaches compared to the deconvolution techniques in \citet{Farrens2017} in simulations of realistic galaxy images derived from HST observations, with realistic sampled sparse-variant PSFs and noise, processed with the GalSim simulation code. We investigated in particular how to set the hyperparameters in both approach: the Tikhonov hyperparameter for the Tikhonet and the continuation parameters for the ADMMnet and compared our proposed XDense U-Net architecture with a "classical" U-Net implementation.
Our main findings are as follows:
\begin{itemize}
\item for both Tikhonet and ADMMnet, the hyperparameters impact the performance of the approaches, but the results are quite stable in a range of values for these hyperparameters. In particular for the Tikhonet, the SURE minimizer is within this range. For the ADMMnet, more hyperparameters needs to be set, and the initialization of the augmented lagrangian parameter impacts  the performance: small parameters lead to higher frequencies in the images, while larger parameters lead to over-smooth galaxies recovered.
\item compared to the "classical" implementation, the XDense U-Net leads to consistently improved criteria for the Tikhonet approach; the situation is more balanced for the ADMMnet, where lower pixel errors can be achieved at low SNR with the "classical" architecture (with high hyperparamer value), but the XDense U-Net provides the best results for pixel errors at high SNR and ellipticity errors both at high and low SNR (with low hyperparameter value); however selecting one or the other architecture with their best hyperparameter value would not change the ranking among methods
\item visually both methods outperform the sparse recovery and low-rank techniques, which displays artefacts at the low SNR probed (and for high SNR as well in the low-rank approach)
\item this is also confirmed in all SNR ranges and for a realistic distribution of SNR; in the latter about $14\%$ improvement is achieved in terms of median pixel error and about $13\%$ improvement for median shape measurement errors for the Tikhonet compared to sparse recovery.
\item among DNN approaches, Tikhonet outperforms ADMMnet in terms of median pixel errors whatever the SNR, and median ellipticity errors for low SNR ($\SNR<40$). At higher SNR, the ADMMnet leads to slightly lower ellipticity errors.
\item the Tikhonet is the fastest approach once the network parameters have been learnt, with about 0.05s needed to process a galaxy, to be compared with sparse and ADMMnet iterative deconvolution approaches which takes about 7 to 10s per galaxy.
\end{itemize}

If the ADMMnet approach is still promising, as extra constraints could be added easily to the framework (while the success of the Tikhonet approach also lies on the ability to compute a closed-form solution for the deconvolution step), these results illustrate that the Tikhonet is overall the best approach in this scenario to process both with high accuracy and fastly a large number of galaxies.

\section*{Reproducible Research} 

In the spirit of reproducible research, the codes will be made freely available on the CosmoStat website. The testing datasets will also be provided to repeat the experiments performed in this paper.

\begin{acknowledgements}
  The authors thank the Galsim developers/GREAT3 collaboration for publicly providing simulation codes and galaxy databases, and the developers of sf\_deconvolve and shapelens as well for their code publicly available.
\end{acknowledgements}

\bibliographystyle{aa} 
\bibliography{references} 

\begin{thebibliography}{94}
\expandafter\ifx\csname natexlab\endcsname\relax\def\natexlab#1{#1}\fi

\bibitem[{Adler \& {\"O}ktem(2017)}]{Adler2017}
Adler, J. \& {\"O}ktem, O. 2017, Inverse Problems, 33, 124007

\bibitem[{{Adler} \& {{\"O}ktem}(2018)}]{Adler2018}
{Adler}, J. \& {{\"O}ktem}, O. 2018, IEEE Transactions on Medical Imaging, 37,
  1322

\bibitem[{Afonso {et~al.}(2011)Afonso, Bioucas-Dias, \&
  Figueiredo}]{Afonso2011}
Afonso, M., Bioucas-Dias, J., \& Figueiredo, M. 2011, IEEE transactions on
  image processing, 20, 681

\bibitem[{Andrews \& Hunt(1977)}]{Andrews77}
Andrews, H.~C. \& Hunt, B.~R. 1977, Digital Image Restoration (Englewood
  Cliffs, NJ: Prentice-Hall)

\bibitem[{Beck \& Teboulle(2009)}]{Beck2009}
Beck, A. \& Teboulle, M. 2009, SIAM Journal on Imaging Sciences, 2, 183

\bibitem[{Bertero \& Boccacci(1998)}]{ima:bertero98}
Bertero, M. \& Boccacci, P. 1998, Introduction to Inverse Problems in Imaging
  (Institute of Physics)

\bibitem[{{Bertin}(2011)}]{Bertin2011}
{Bertin}, E. 2011, in Astronomical Society of the Pacific Conference Series,
  Vol. 442, Astronomical Data Analysis Software and Systems XX, ed. I.~N.
  {Evans}, A.~{Accomazzi}, D.~J. {Mink}, \& A.~H. {Rots}, 435

\bibitem[{{Bertocchi} {et~al.}(2018){Bertocchi}, {Chouzenoux}, {Corbineau},
  {Pesquet}, \& {Prato}}]{Bertocchi2019}
{Bertocchi}, C., {Chouzenoux}, E., {Corbineau}, M.-C., {Pesquet}, J.-C., \&
  {Prato}, M. 2018, arXiv e-prints, arXiv:1812.04276

\bibitem[{Bigdeli {et~al.}(2017)Bigdeli, Jin, Favaro, \& Zwicker}]{Bigdeli2017}
Bigdeli, S.~A., Jin, M., Favaro, P., \& Zwicker, M. 2017, in Proceedings of the
  31st International Conference on Neural Information Processing Systems,
  NIPS'17 (USA: Curran Associates Inc.), 763--772

\bibitem[{Bioucas-Dias(2006)}]{BioucasDias2006}
Bioucas-Dias, J. 2006, Image Processing, IEEE Transactions on, 15, 937

\bibitem[{{Bobin} {et~al.}(2014){Bobin}, {Sureau}, {Starck}, {Rassat}, \&
  {Paykari}}]{PR1_WPR1}
{Bobin}, J., {Sureau}, F., {Starck}, J.-L., {Rassat}, A., \& {Paykari}, P.
  2014, \aap, 563, A105

\bibitem[{Boyd {et~al.}(2010)Boyd, Parikh, Chu, Peleato, \&
  Eckstein}]{Boyd2010}
Boyd, S., Parikh, N., Chu, E., Peleato, B., \& Eckstein, J. 2010, Machine
  Learning, 3, 1

\bibitem[{C.~Eldar(2009)}]{Eldar2009}
C.~Eldar, Y. 2009, Signal Processing, IEEE Transactions on, 57, 471

\bibitem[{Cai {et~al.}(2010)Cai, Osher, \& Shen}]{Cai2010}
Cai, J., Osher, S., \& Shen, Z. 2010, Multiscale Modeling \& Simulation, 8, 337

\bibitem[{Chambolle \& Pock(2011)}]{Chambolle2011}
Chambolle, A. \& Pock, T. 2011, Journal of Mathematical Imaging and Vision, 40,
  120

\bibitem[{Chollet(2016)}]{Chollet2016}
Chollet, F. 2016, 2017 IEEE Conference on Computer Vision and Pattern
  Recognition (CVPR), 1800

\bibitem[{Christian~Hansen \& O'leary(1993)}]{Hansen1993}
Christian~Hansen, P. \& O'leary, D. 1993, SIAM J. Sci. Comput., 14, 1487

\bibitem[{Combettes \& Pesquet(2011)}]{Combettes2011}
Combettes, P.~L. \& Pesquet, J.-C. 2011, in {Fixed-Point Algorithms for Inverse
  Problems in Science and Engineering}, ed. Bauschke, H.~Burachik,
  R.~Combettes, P.~Elser, V.~Luke, D.~Wolkowicz, \& H.~(Eds.) ({Springer}),
  185--212

\bibitem[{Combettes \& Vu(2014)}]{Combettes2014}
Combettes, P.~L. \& Vu, B.~C. 2014, Optimization, 63, 1289

\bibitem[{Condat(2013)}]{Condat2013}
Condat, L. 2013, {Journal of Optimization Theory and Applications}, 158, 460

\bibitem[{Deledalle {et~al.}(2014)Deledalle, Vaiter, Fadili, \&
  Peyr{\'e}}]{Deledalle2014}
Deledalle, C., Vaiter, S., Fadili, J., \& Peyr{\'e}, G. 2014, SIAM Journal on
  Imaging Sciences, 7, 2448

\bibitem[{Donoho(1995)}]{Donoho1995}
Donoho, D.~L. 1995, Applied and Computational Harmonic Analysis, 2, 101

\bibitem[{Eldan \& Shamir(2015)}]{Eldan2015}
Eldan, R. \& Shamir, O. 2015, Conference on Learning Theory

\bibitem[{Elfwing {et~al.}(2018)Elfwing, Uchibe, \& Doya}]{Elfwing2018}
Elfwing, S., Uchibe, E., \& Doya, K. 2018, Neural Networks, 107, 3 , special
  issue on deep reinforcement learning

\bibitem[{Fan {et~al.}(2018)Fan, Li, Teng, \& Wang}]{fan2018soft}
Fan, F., Li, M., Teng, Y., \& Wang, G. 2018, arXiv preprint arXiv:1812.11675

\bibitem[{Farrens {et~al.}(2017)Farrens, Starck, \& Mboula}]{Farrens2017}
Farrens, S., Starck, J.-L., \& Mboula, F. 2017, Astronomy \& Astrophysics, 601

\bibitem[{Flamary(2017)}]{Flamary2017}
Flamary, R. 2017, 2017 25th European Signal Processing Conference (EUSIPCO),
  2468

\bibitem[{{Garsden} {et~al.}(2015){Garsden}, {Girard}, \&
  {Starck}}]{starck:garsden2015}
{Garsden}, H., {Girard}, J.~N., \& {Starck}, J.-L., e.~a. 2015, \aa, 575, A90

\bibitem[{Golub {et~al.}(1979)Golub, Heath, \& Wahba}]{Golub1979}
Golub, G.~H., Heath, M., \& Wahba, G. 1979, Technometrics, 21, 215

\bibitem[{Gregor \& LeCun(2010)}]{Gregor2010}
Gregor, K. \& LeCun, Y. 2010, in Proceedings of the 27th International
  Conference on International Conference on Machine Learning, ICML'10 (USA:
  Omnipress), 399--406

\bibitem[{{Guerrero-colon} \& {Portilla}(2006)}]{Guerrero2006}
{Guerrero-colon}, J.~A. \& {Portilla}, J. 2006, in 2006 International
  Conference on Image Processing, 625--628

\bibitem[{{Gupta} {et~al.}(2018){Gupta}, {Jin}, {Nguyen}, {McCann}, \&
  {Unser}}]{Gupta2018}
{Gupta}, H., {Jin}, K.~H., {Nguyen}, H.~Q., {McCann}, M.~T., \& {Unser}, M.
  2018, IEEE Transactions on Medical Imaging, 37, 1440

\bibitem[{H.~Chan {et~al.}(2016)H.~Chan, Wang, \& A.~Elgendy}]{Chan2016}
H.~Chan, S., Wang, X., \& A.~Elgendy, O. 2016, IEEE Transactions on
  Computational Imaging, PP

\bibitem[{Han \& Ye(2018)}]{han2018framing}
Han, Y. \& Ye, J.~C. 2018, IEEE transactions on medical imaging, 37, 1418

\bibitem[{{He} {et~al.}(2016){He}, {Zhang}, {Ren}, \& {Sun}}]{He2016}
{He}, K., {Zhang}, X., {Ren}, S., \& {Sun}, J. 2016, in 2016 IEEE Conference on
  Computer Vision and Pattern Recognition (CVPR), 770--778

\bibitem[{{Huang} {et~al.}(2017){Huang}, {Liu}, v.~d. {Maaten}, \&
  {Weinberger}}]{Huang2017}
{Huang}, G., {Liu}, Z., v.~d. {Maaten}, L., \& {Weinberger}, K.~Q. 2017, in
  2017 IEEE Conference on Computer Vision and Pattern Recognition (CVPR),
  2261--2269

\bibitem[{Hunt(1972)}]{Hunt1972}
Hunt, B.~R. 1972, {IEEE} {T}ransactions on {A}utomatic and {C}ontrol, AC-17,
  703

\bibitem[{Ioffe \& Szegedy(2015)}]{Ioffe2015}
Ioffe, S. \& Szegedy, C. 2015, in Proceedings of the 32Nd International
  Conference on International Conference on Machine Learning - Volume 37,
  ICML'15 (JMLR.org), 448--456

\bibitem[{{Jia} \& {Evans}(2011)}]{Jia2011}
{Jia}, C. \& {Evans}, B.~L. 2011, in 2011 18th IEEE International Conference on
  Image Processing, 681--684

\bibitem[{{Jin} {et~al.}(2017){Jin}, {McCann}, {Froustey}, \&
  {Unser}}]{Jin2017}
{Jin}, K.~H., {McCann}, M.~T., {Froustey}, E., \& {Unser}, M. 2017, IEEE
  Transactions on Image Processing, 26, 4509

\bibitem[{{Kaiser} {et~al.}(1995){Kaiser}, {Squires}, \&
  {Broadhurst}}]{Kaiser1995}
{Kaiser}, N., {Squires}, G., \& {Broadhurst}, T. 1995, \apj, 449, 460

\bibitem[{{Kalifa} {et~al.}(2003){Kalifa}, {Mallat}, \& {Rouge}}]{Kalifa2003}
{Kalifa}, J., {Mallat}, S., \& {Rouge}, B. 2003, IEEE Transactions on Image
  Processing, 12, 446

\bibitem[{Kingma \& Ba(2014)}]{Kingma2014}
Kingma, D. \& Ba, J. 2014, International Conference on Learning Representations

\bibitem[{Krishnan \& Fergus(2009)}]{Krishnan2009}
Krishnan, D. \& Fergus, R. 2009, in Advances in Neural Information Processing
  Systems 22, ed. Y.~Bengio, D.~Schuurmans, J.~D. Lafferty, C.~K.~I. Williams,
  \& A.~Culotta (Curran Associates, Inc.), 1033--1041

\bibitem[{{Krist} {et~al.}(2011){Krist}, {Hook}, \& {Stoehr}}]{Krist2011}
{Krist}, J.~E., {Hook}, R.~N., \& {Stoehr}, F. 2011, in \procspie, Vol. 8127,
  Optical Modeling and Performance Predictions V, 81270J

\bibitem[{{Kuijken} {et~al.}(2015){Kuijken}, {Heymans}, {Hildebrandt},
  {Nakajima}, {Erben}, {de Jong}, {Viola}, {Choi}, {Hoekstra}, {Miller}, {van
  Uitert}, {Amon}, {Blake}, {Brouwer}, {Buddendiek}, {Conti}, {Eriksen},
  {Grado}, {Harnois-D{\'e}raps}, {Helmich}, {Herbonnet}, {Irisarri},
  {Kitching}, {Klaes}, {La Barbera}, {Napolitano}, {Radovich}, {Schneider},
  {Sif{\'o}n}, {Sikkema}, {Simon}, {Tudorica}, {Valentijn}, {Verdoes Kleijn},
  \& {van Waerbeke}}]{Kuijken2015}
{Kuijken}, K., {Heymans}, C., {Hildebrandt}, H., {et~al.} 2015, \mnras, 454,
  3500

\bibitem[{{Lanusse, F.} {et~al.}(2016){Lanusse, F.}, {Starck, J.-L.}, {Leonard,
  A.}, \& {Pires, S.}}]{glimpse2016}
{Lanusse, F.}, {Starck, J.-L.}, {Leonard, A.}, \& {Pires, S.} 2016, A\&A, 591,
  A2

\bibitem[{LeCun {et~al.}(2015)LeCun, Bengio, \& Hinton}]{Lecun2015}
LeCun, Y., Bengio, Y., \& Hinton, G. 2015, Nature, 521, 436

\bibitem[{{Li} {et~al.}(2019){Li}, {Tofighi}, {Monga}, \& {Eldar}}]{Li2019}
{Li}, Y., {Tofighi}, M., {Monga}, V., \& {Eldar}, Y.~C. 2019, in ICASSP 2019 -
  2019 IEEE International Conference on Acoustics, Speech and Signal Processing
  (ICASSP), 7675--7679

\bibitem[{Lou {et~al.}(2011)Lou, Bertozzi, \& Soatto}]{Lou2011}
Lou, Y., Bertozzi, A.~L., \& Soatto, S. 2011, Journal of Mathematical Imaging
  and Vision, 39, 1

\bibitem[{Mairal {et~al.}(2008)Mairal, Sapiro, \& Elad}]{Mairal2008}
Mairal, J., Sapiro, G., \& Elad, M. 2008, Multiscale Modeling \& Simulation, 7,
  214

\bibitem[{Mallat(2016)}]{Mallat2016}
Mallat, S. 2016, Philosophical Transactions of the Royal Society A:
  Mathematical, Physical and Engineering Sciences, 374, 20150203

\bibitem[{{Mandelbaum} {et~al.}(2015){Mandelbaum}, {Rowe}, {Armstrong}, {Bard},
  {Bertin}, {Bosch}, {Boutigny}, {Courbin}, {Dawson}, {Donnarumma}, {Fenech
  Conti}, {Gavazzi}, {Gentile}, {Gill}, {Hogg}, {Huff}, {Jee}, {Kacprzak},
  {Kilbinger}, {Kuntzer}, {Lang}, {Luo}, {March}, {Marshall}, {Meyers},
  {Miller}, {Miyatake}, {Nakajima}, {Ngol{\'e} Mboula}, {Nurbaeva}, {Okura},
  {Paulin-Henriksson}, {Rhodes}, {Schneider}, {Shan}, {Sheldon}, {Simet},
  {Starck}, {Sureau}, {Tewes}, {Zarb Adami}, {Zhang}, \&
  {Zuntz}}]{Mandelbaum2015}
{Mandelbaum}, R., {Rowe}, B., {Armstrong}, R., {et~al.} 2015, \mnras, 450, 2963

\bibitem[{Mandelbaum {et~al.}(2014)Mandelbaum, Rowe, Bosch, Chang, Courbin,
  Gill, Jarvis, Kannawadi, Kacprzak, Lackner, Leauthaud, Miyatake, Nakajima,
  Rhodes, Simet, Zuntz, Armstrong, Bridle, Coupon, Dietrich, Gentile, Heymans,
  Jurling, Kent, Kirkby, Margala, Massey, Melchior, Peterson, Roodman, \&
  Schrabback}]{Mandelbaum2014}
Mandelbaum, R., Rowe, B., Bosch, J., {et~al.} 2014, The Astrophysical Journal
  Supplement Series, 212, 5

\bibitem[{Mardani {et~al.}(2017)Mardani, Gong, Cheng, Pauly, \&
  Xing}]{Mardani2017}
Mardani, M., Gong, E., Cheng, J.~Y., Pauly, J.~M., \& Xing, L. 2017, in 2017
  {IEEE} 7th International Workshop on Computational Advances in Multi-Sensor
  Adaptive Processing, {CAMSAP} 2017, Cura{\c{c}}ao, The Netherlands, December
  10-13, 2017, 1--5

\bibitem[{Mardani {et~al.}(2018)Mardani, Sun, Vasawanala, Papyan, Monajemi,
  Pauly, \& Donoho}]{Mardani2018}
Mardani, M., Sun, Q., Vasawanala, S., {et~al.} 2018, in Proceedings of the 32Nd
  International Conference on Neural Information Processing Systems, NIPS'18
  (USA: Curran Associates Inc.), 9596--9606

\bibitem[{Mboula {et~al.}(2016)Mboula, Starck, Okumura, Amiaux, \&
  Hudelot}]{Mboula2016}
Mboula, F., Starck, J.-L., Okumura, K., Amiaux, J., \& Hudelot, P. 2016,
  Inverse Problems, 32

\bibitem[{Meinhardt {et~al.}(2017)Meinhardt, Moller, Hazirbas, \&
  Cremers}]{Meinhardt2017}
Meinhardt, T., Moller, M., Hazirbas, C., \& Cremers, D. 2017, in Proceedings of
  the IEEE International Conference on Computer Vision, 1781--1790

\bibitem[{Monga {et~al.}(2019)Monga, Li, \& Eldar}]{Monga2019}
Monga, V., Li, Y., \& Eldar, Y.~C. 2019, Algorithm Unrolling: Interpretable,
  Efficient Deep Learning for Signal and Image Processing

\bibitem[{{Neelamani} {et~al.}(2004){Neelamani}, {Hyeokho Choi}, \&
  {Baraniuk}}]{Neelamani2004}
{Neelamani}, R., {Hyeokho Choi}, \& {Baraniuk}, R. 2004, IEEE Transactions on
  Signal Processing, 52, 418

\bibitem[{Oliveira {et~al.}(2009)Oliveira, Bioucas-Dias, \&
  Figueiredo}]{Oliveira2009}
Oliveira, J.~P., Bioucas-Dias, J.~M., \& Figueiredo, M.~A. 2009, Signal
  Processing, 89, 1683

\bibitem[{{Orieux} {et~al.}(2010){Orieux}, {Giovannelli}, \&
  {Rodet}}]{Orieux2010}
{Orieux}, F., {Giovannelli}, J., \& {Rodet}, T. 2010, in 2010 IEEE
  International Conference on Acoustics, Speech and Signal Processing,
  1350--1353

\bibitem[{{Pereyra} {et~al.}(2015){Pereyra}, {Bioucas-Dias}, \&
  {Figueiredo}}]{Pereyra2015}
{Pereyra}, M., {Bioucas-Dias}, J.~M., \& {Figueiredo}, M. A.~T. 2015, in 2015
  23rd European Signal Processing Conference (EUSIPCO), 230--234

\bibitem[{{Pesquet} {et~al.}(2009){Pesquet}, {Benazza-Benyahia}, \&
  {Chaux}}]{Pesquet2009}
{Pesquet}, J., {Benazza-Benyahia}, A., \& {Chaux}, C. 2009, IEEE Transactions
  on Signal Processing, 57, 4616

\bibitem[{Petersen \& Voigtlaender(2018)}]{Petersen2018}
Petersen, P. \& Voigtlaender, F. 2018, Neural Networks, 108, 296

\bibitem[{Pustelnik {et~al.}(2016)Pustelnik, Benazza-Benhayia, Zheng, \&
  Pesquet}]{Pustelnik2016}
Pustelnik, N., Benazza-Benhayia, A., Zheng, Y., \& Pesquet, J.-C. 2016, {Wiley
  Encyclopedia of Electrical and Electronics Engineering}

\bibitem[{{Ribli} {et~al.}(2019){Ribli}, {Pataki}, \& {Csabai}}]{Ribli2018}
{Ribli}, D., {Pataki}, B.~{\'A}., \& {Csabai}, I. 2019, Nature Astronomy, 3, 93

\bibitem[{Romano {et~al.}(2017)Romano, Elad, \& Milanfar}]{Romano2017}
Romano, Y., Elad, M., \& Milanfar, P. 2017, SIAM Journal on Imaging Sciences,
  10, 1804

\bibitem[{Ronneberger {et~al.}(2015)Ronneberger, Fischer, \&
  Brox}]{Ronneberger2015}
Ronneberger, O., Fischer, P., \& Brox, T. 2015, in Medical Image Computing and
  Computer-Assisted Intervention -- MICCAI 2015, ed. N.~Navab, J.~Hornegger,
  W.~M. Wells, \& A.~F. Frangi (Cham: Springer International Publishing),
  234--241

\bibitem[{Rowe {et~al.}(2015)Rowe, Jarvis, Mandelbaum, Bernstein, Bosch, Simet,
  Meyers, Kacprzak, Nakajima, Zuntz, Miyatake, Dietrich, Armstrong, Melchior,
  \& Gill}]{Rowe2015}
Rowe, B., Jarvis, M., Mandelbaum, R., {et~al.} 2015, Astronomy and Computing,
  10, 121

\bibitem[{Safran \& Shamir(2017)}]{Safran2017}
Safran, I. \& Shamir, O. 2017, in Proceedings of the 34th International
  Conference on Machine Learning - Volume 70, ICML'17 (JMLR.org), 2979--2987

\bibitem[{Schawinski {et~al.}(2017)Schawinski, Zhang, Zhang, Fowler, \&
  Santhanam}]{Schawinski2017}
Schawinski, K., Zhang, C., Zhang, H., Fowler, L., \& Santhanam, G.~K. 2017,
  Monthly Notices of the Royal Astronomical Society: Letters, 467, slx008

\bibitem[{{Schmitz} {et~al.}(2019){Schmitz}, {Starck}, {Ngole Mboula},
  {Auricchio}, {Brinchmann}, {Vito Capobianco}, {Cl{\'e}dassou}, {Conversi},
  {Corcione}, {Fourmanoit}, {Frailis}, {Garilli}, {Hormuth}, {Hu}, {Israel},
  {Kermiche}, {Kitching}, {Kubik}, {Kunz}, {Ligori}, {Lilje}, {Lloro},
  {Mansutti}, {Marggraf}, {Massey}, {Pasian}, {Pettorino}, {Raison}, {Rhodes},
  {Roncarelli}, {Saglia}, {Schneider}, {Serrano}, {Taylor}, {Toledo-Moreo},
  {Valenziano}, {Vuerli}, \& {Zoubian}}]{Schmitz2019}
{Schmitz}, M.~A., {Starck}, J.~L., {Ngole Mboula}, F., {et~al.} 2019, arXiv
  e-prints, arXiv:1906.07676

\bibitem[{{Schuler} {et~al.}(2013){Schuler}, {Burger}, {Harmeling}, \&
  {Sch{\"o}lkopf}}]{Schuler2013}
{Schuler}, C.~J., {Burger}, H.~C., {Harmeling}, S., \& {Sch{\"o}lkopf}, B.
  2013, in 2013 IEEE Conference on Computer Vision and Pattern Recognition,
  1067--1074

\bibitem[{{Schuler} {et~al.}(2016){Schuler}, {Hirsch}, {Harmeling}, \&
  {Sch{\"o}lkopf}}]{Schuler2016}
{Schuler}, C.~J., {Hirsch}, M., {Harmeling}, S., \& {Sch{\"o}lkopf}, B. 2016,
  IEEE Transactions on Pattern Analysis and Machine Intelligence, 38, 1439

\bibitem[{Sreehari {et~al.}(2016)Sreehari, Venkatakrishnan, Wohlberg, Buzzard,
  Drummy, Simmons, \& Bouman}]{Sreehari2016}
Sreehari, S., Venkatakrishnan, S., Wohlberg, B., {et~al.} 2016, IEEE
  Transactions on Computational Imaging, 2, 408

\bibitem[{Starck {et~al.}(2000)Starck, Bijaoui, Valtchanov, \&
  Murtagh}]{starck:sta00_3}
Starck, J.-L., Bijaoui, A., Valtchanov, I., \& Murtagh, F. 2000, Astronomy and
  Astrophysics, Supplement Series, 147, 139--149

\bibitem[{Starck {et~al.}(2015{\natexlab{a}})Starck, Murtagh, \&
  Bertero}]{Starck2015}
Starck, J.-L., Murtagh, F., \& Bertero, M. 2015{\natexlab{a}}, Starlet
  Transform in Astronomical Data Processing (New York, NY: Springer New York),
  2053--2098

\bibitem[{Starck {et~al.}(2015{\natexlab{b}})Starck, Murtagh, \&
  Fadili}]{starck:book15}
Starck, J.-L., Murtagh, F., \& Fadili, J. 2015{\natexlab{b}}, Sparse Image and
  Signal Processing: Wavelets and Related Geometric Multiscale Analysis
  (Cambridge University Press)

\bibitem[{Starck {et~al.}(2003)Starck, Nguyen, \& Murtagh}]{Starck2003}
Starck, J.-L., Nguyen, M., \& Murtagh, F. 2003, Signal Processing, 83, 2279

\bibitem[{Sureau {et~al.}(2014)Sureau, Starck, Bobin, Paykari, \&
  Rassat}]{Sureau2014}
Sureau, F.~C., Starck, J.-L., Bobin, J., Paykari, P., \& Rassat, A. 2014,
  {Astronomy and Astrophysics - A\&A}, 566, A100

\bibitem[{Szegedy {et~al.}(2016)Szegedy, Ioffe, \& Vanhoucke}]{Szegedy2016}
Szegedy, C., Ioffe, S., \& Vanhoucke, V. 2016, AAAI Conference on Artificial
  Intelligence

\bibitem[{T.~Reehorst \& Schniter(2018)}]{Reehorst2018}
T.~Reehorst, E. \& Schniter, P. 2018, IEEE Transactions on Computational
  Imaging, PP, 1

\bibitem[{Tikhonov \& Arsenin(1977)}]{Tikhonov1977}
Tikhonov, A.~N. \& Arsenin, V.~Y. 1977, Solutions of Ill-posed problems
  (W.H.~Winston)

\bibitem[{Twomey(1963)}]{Twomey1963}
Twomey, S. 1963, J. ACM, 10, 97

\bibitem[{Venkatakrishnan {et~al.}(2013)Venkatakrishnan, Bouman, \&
  Wohlberg}]{Venkatakrishnan2013}
Venkatakrishnan, S.~V., Bouman, C.~A., \& Wohlberg, B. 2013, 2013 IEEE Global
  Conference on Signal and Information Processing, 945

\bibitem[{{Viola} {et~al.}(2011){Viola}, {Melchior}, \&
  {Bartelmann}}]{Viola2011}
{Viola}, M., {Melchior}, P., \& {Bartelmann}, M. 2011, \mnras, 410, 2156

\bibitem[{Werner~Engl {et~al.}(1996)Werner~Engl, Hanke, \& Neubauer}]{Engl1996}
Werner~Engl, H., Hanke, M., \& Neubauer, A. 1996, Regularization of inverse
  problems (Dordrecht:Kluwer)

\bibitem[{Xu {et~al.}(2014)Xu, Ren, Liu, \& Jia}]{Xu2014}
Xu, L., Ren, J., Liu, C., \& Jia, J. 2014, Advances in Neural Information
  Processing Systems, 2, 1790

\bibitem[{Ye {et~al.}(2018{\natexlab{a}})Ye, Han, \& Cha}]{Ye2018}
Ye, J., Han, Y., \& Cha, E. 2018{\natexlab{a}}, SIAM Journal on Imaging
  Sciences, 11, 991

\bibitem[{Ye {et~al.}(2018{\natexlab{b}})Ye, Han, \& Cha}]{ye2018deep}
Ye, J.~C., Han, Y., \& Cha, E. 2018{\natexlab{b}}, SIAM Journal on Imaging
  Sciences, 11, 991

\bibitem[{{Zhang} {et~al.}(2017){Zhang}, {Zuo}, {Gu}, \& {Zhang}}]{Zhang2017}
{Zhang}, K., {Zuo}, W., {Gu}, S., \& {Zhang}, L. 2017, in 2017 IEEE Conference
  on Computer Vision and Pattern Recognition (CVPR), 2808--2817

\bibitem[{{Zibulevsky} \& {Elad}(2010)}]{Zibulevsky2010}
{Zibulevsky}, M. \& {Elad}, M. 2010, IEEE Signal Processing Magazine, 27, 76

\bibitem[{Zuntz {et~al.}(2018)Zuntz, Sheldon, Samuroff, Troxel, Jarvis,
  MacCrann, Gruen, Prat, S{\'a}nchez, Choi, Bridle, Bernstein, Dodelson,
  Drlica-Wagner, Fang, Gruendl, Hoyle, Huff, Jain, Kirk, Kacprzak, Krawiec,
  Plazas, Rollins, Rykoff, Sevilla-Noarbe, Soergel, Varga, Abbott, Abdalla,
  Allam, Annis, Bechtol, Benoit-L{\'e}vy, Bertin, Buckley-Geer, Burke,
  Carnero~Rosell, Kind, Carretero, Castander, Crocce, Cunha, D'Andrea,
  da~Costa, Davis, Desai, Diehl, Dietrich, Doel, Eifler, Estrada, Evrard, Neto,
  Fernandez, Flaugher, Fosalba, Frieman, Garc{\'\i}a-Bellido, Gaztanaga,
  Gerdes, Giannantonio, Gschwend, Gutierrez, Hartley, Honscheid, James,
  Jeltema, Johnson, Johnson, Kuehn, Kuhlmann, Kuropatkin, Lahav, Li, Lima,
  Maia, March, Martini, Melchior, Menanteau, Miller, Miquel, Mohr, Neilsen,
  Nichol, Ogando, Roe, Romer, Roodman, Sanchez, Scarpine, Schindler, Schubnell,
  Smith, Smith, Soares-Santos, Sobreira, Suchyta, Swanson, Tarle, Thomas,
  Tucker, Vikram, Walker, Wechsler, Zhang, \& Collaboration)}]{Zuntz2018}
Zuntz, J., Sheldon, E., Samuroff, S., {et~al.} 2018, Monthly Notices of the
  Royal Astronomical Society, 481, 1149

\end{thebibliography}
\end{document}